\documentclass{aa}  
\usepackage{natbib}
\usepackage{graphicx}
\usepackage{txfonts}
\usepackage{multirow}
\usepackage{lipsum}
\usepackage{subcaption}
\usepackage[dvipsnames]{xcolor}
\PassOptionsToPackage{hyphens}{url}
\usepackage[colorlinks, citecolor=blue]{hyperref}
\usepackage{cancel}
\usepackage{soul}

\titlerunning{Searching for compact objects in SB1 systems in NGC\,6231}
\authorrunning{G.\ Banyard et al.}

\newcommand{\kms}{km\,${\rm s}^{-1}$}

\begin{document} 
  \title{Searching for compact objects in the single-lined spectroscopic binaries of the young Galactic cluster NGC 6231\thanks{Based on observations collected at the ESO Paranal observatory under ESO program 099.D-0895 and 0101.D-0163.}}
  \author{G. Banyard\inst{1} %\orcid{0000-0002-6374-9903}
     \and
          L. Mahy\inst{2}\fnmsep
     \and
          H. Sana\inst{1}
     \and
%          F. Riends
          J. Bodensteiner\inst{3}
     \and 
          J.~I. Villase\~nor\inst{1} %\orcid{0000-0002-7984-1675}
     \and 
          K. Sen\inst{5}
     \and 
          N. Langer\inst{4,5}
     \and 
          S. de Mink\inst{6,7}
     \and 
          A. Picco\inst{1}
     \and 
          T. Shenar\inst{7}
%           \thanks{Just to show the usage of the elements in the author field}
          }

  \institute{Institute for Astronomy, KU Leuven,
             Celestijnenlaan 200 D, 3001, Leuven, Belgium
             \and 
             Royal Observatory of Belgium, Avenue Circulaire/Ringlaan 3, B-1180 Brussels, Belgium
             \and
             ESO – European Organisation for Astronomical Research in the Southern Hemisphere, Karl-Schwarzschild-Strasse 2, 85748 Garching, Germany
             \and 
             Argelander-Institut f{\"u}r Astronomie, Universit{\"a}t Bonn, Auf dem H{\"u}gel 71, 53121 Bonn, Germany
             \and 
             Max-Planck-Institut f{\"u}r Radioastronomie, Auf dem H{\"u}gel 69, 53121 Bonn, Germany
             \and
             Max-Planck Institute for Astrophysics, Karl-Schwarzschild-Str. 1, 85748 Garching, Germany
             \and
             Anton Pannekoek Institute for Astronomy, University of Amsterdam, Postbus 94249, 1090 GE Amsterdam, The Netherlands
             }
  %           \and
  %           Institute for Astronomy, University of Edinburgh, Royal Observatory, Blackford Hill, Edinburgh, EH9 3HJ, UK
   %          \and 
    %         UK Astronomy Technology Centre, Royal Observatory, Blackford Hill, Edinburgh, EH9 3HJ}

  \date{}

  \abstract
  % context heading (optional)
  % {} leave it empty if necessary
   {Recent evolutionary computations predict that a few percent of massive O- or early-B stars in binary systems should have a dormant BH as a companion. However, despite several reported candidate X-ray quiet OB+BH systems over the last couple of years, finding them with certainty remains challenging. Yet these have great importance as they as they can be gravitational wave (GW) source progenitors, and are landmark systems in constraining supernova kick physics.}
  % aims heading (mandatory)
   {This work aims to characterise the hidden companions to the single-lined spectroscopic binaries (SB1s) identified in the B star population of the young open Galactic cluster NGC 6231 to find candidate systems for harbouring compact object companions.}
  % methods heading (mandatory)
   {With the orbital solutions for each SB1 constrained in a previous study, we applied Fourier spectral disentangling to multi-epoch optical VLT/FLAMES spectra of each target to extract a potential signature of a faint companion, and to identify newly disentangled double-lined spectroscopic binaries (SB2s). For targets where the disentangling does not reveal any spectral signature of a stellar companion, we performed atmospheric and evolutionary modelling on the primary (visible) star to obtain constraints on the mass and nature of the unseen companion.}
  % results heading (mandatory)
   {For 7 of the 15 apparent SB1 systems, we extracted the spectral signature of a faint companion, resulting in seven newly classified SB2 systems with mass ratios down to near 0.1. From the remaining targets, for which no faint companion could be extracted from the spectra, four are found to have companion masses that lie in the predicted mass ranges of neutron stars (NSes) and BHes. Two of these targets have companion masses between 1 and 3.5 $M_{\odot}$, making them potential hosts of NSes (or lower mass main sequence stars). The other two have mass ranges between 2.5 to 8 $M_{\odot}$ and 1.6 and 26 $M_{\odot}$, respectively, and so are identified as candidates for harbouring BH companions.}
  % conclusions heading (optional), leave it empty if necessary 
   {We present four SB1 systems in NGC 6231 that are candidates for harbouring compact objects. However, unambiguous identification of these systems as X-ray quiet compact object harbouring binaries will only result from photometric and interferometric follow-up observations of these objects.}


\maketitle

%________________________________________________________________

\section{Introduction}
\label{s:intro}

Massive stars play a fundamental role in many branches of astrophysics during their relatively short lives. This is especially true during their end stages, where their explosions as core collapse supernovae (CCSNe) impart energy and momentum into their surroundings, as well as chemically enrich them with processed material. These cataclysmic events drive and alter the dynamical and chemical evolution of their host galaxies \citep{hopkins_galaxies_2014}, and produce neutron stars (NS) and black holes (BH) as remnant compact objects, depending on the mass of the progenitor star. \par

A complicating factor in the pre-supernova life of a massive star is their prevalence in multiple systems, especially in those with periods short enough to allow for interaction between members. This has been observed in several studies of large, statistically significant populations of massive stars, both spectroscopically and interferometrically \citep{sana_massive_2008,mahy_early-type_2009,sana_massive_2009,sana_massive_2011,evans_vlt_2011,sana_binary_2012,kobulnicky_fresh_2012,kobulnicky_toward_2014,mahy_spectroscopic_2013,sana_southern_2014,barba_own_2017,almeida_tarantula_2017,trigueros_paez_monos_2021,banyard_observed_2022}.
These campaigns aimed to constrain the binary fraction of OB stars in clusters and associations (both in and out of the Galaxy) and to characterise the multiplicity properties of the identified binary systems. One of these works targeted the B stars of NGC 6231 \citep[][hereafter Paper~I]{banyard_observed_2022}, a young (7 Myrs) open cluster in the Sco OB2 association in the Galaxy. Through multi-epoch VLT/FLAMES spectroscopy of the targets, an observed binary fraction of $33\pm5$\% was obtained, increasing to $52\pm8\%$ when corrected for the observational biases. Five double-lined (SB2) spectroscopic binaries and 15 single-lined (SB1) spectroscopic binaries were identified in the sample. \par

 %All of the GW events detected by LIGO thus far are from the inspiral and merging of double compact objects in close binary systems. \par

%%REWORK BASED ON TOMER'S COMMENT:
%"Since you already focused the text on “isolated binary evolution” above, I find it distracting to mention this here again. I would first mention the various ideas (dynamical captures, primodial black holes, isolated binary evolution…), and then talk only on binary processes. "
The high frequency of massive stars in multiple systems, paired with their production of compact objects, makes massive stars in binary systems a main progenitor of gravitational wave (GW) sources observed by ground-based gravitational wave detectors. What is insufficiently constrained then, is the mechanism by which these close double compact systems are formed. Various schemes have been proposed. Firstly, there are those that involve isolated binary (multiple) evolution. Schemes involving isolated multiple evolution include the production of the very first binary BH systems from Population III stars \citep{belczynski_first_2004,kinugawa_possible_2014,inayoshi_formation_2017}, and dynamical processes such as interactions in globular and nuclear clusters \citep{kulkarni_stellar_1993,antonini_secular_2012,rodriguez_black_2019}, in isolated triple systems \citep[e.g,][]{antonini_binary_2017}, along with the formation of compact object binaries in active galactic nuclei (AGN) disks \citep{tagawa_formation_2020}. Chemically homogeneous evolution (CHE) \citep{maeder_evidences_1987,langer_helium_1992} can prevent the transfer of mass between stars in very close binaries \citep{de_mink_rotational_2009,marchant_new_2016,de_mink_chemically_2016,abdul-masih_clues_2019,du_buisson_cosmic_2020,riley_chemically_2021,abdul-masih_constraining_2021,menon_detailed_2021}, and this allows for both of the massive stars in a system to directly evolve into BHs. This scenario, however, favours low-metallicity (SMC or lower) and relatively massive ($>30 M_{\odot}$) stars, and so this channel is unlikely to be relevant for Galactic B stars. Alternatively, there are also schemes that involve binary processes. For example, mass transfer between stars via Roche Lobe overflow (RLOF) can result in a common envelope (CE) phase. If the separation is large enough, members of massive binaries can evolve and expand individually before the orbit shrinks due to the CE phase, avoiding premature merging and producing a double compact object system \citep{belczynski_first_2016,langer_properties_2020}. However, the underlying physical processes are not fully understood and the uncertainties in CE evolution are large. Other pathways involving binary processes include stable mass transfer \citep{van_den_heuvel_forming_2017,neijssel_effect_2019,bavera_origin_2020,marchant_role_2021,sen_detailed_2022}. Observing and characterising these GW progenitors, then, will help reduce the uncertainty in their formation scenarios. \par 

Hundreds of millions of stellar mass BHs are predicted to occupy the Galaxy \citep{brown_scenario_1994}, with the nearest of them expected to occupy space within the order of 10 pc \citep{shapiro_book-review_1983,chisholm_stellar-mass_2003}. Recent theoretical computations have predicted that 3\% of massive O- or early-B stars in binary systems should have a dormant stellar-mass BH as a companion \citep{shao_population_2019,langer_properties_2020}. However, observational evidence is still lacking. Apart from the 50 significant GW detections of double BH systems, the majority of BHs (and NSes) are found through X-ray observations, specifically as accretors in X-ray binaries (XRBs). These are detected as such as the compact objects in these systems are accrete material from their companions, either through RLOF episodes or wind accretion. In particular, the population of BH candidates in X-ray binaries has increased considerably in the last 50 years with 59 Galactic objects detected in transient low-mass XRBs alone \citep{corral-santana_blackcat_2016}. The only high-mass X-ray binary (HMXRB) detected thus far in the Galaxy, that is a BH accreting from a massive companion, is Cygnus X-1 \citep{orosz_mass_2011,miller-jones_cygnus_2021}, which hosts a 21 $M_{\odot}$ BH.\par

There are very few X-ray "quiet" massive stars claimed to harbour a BH which have not yet been refuted in the literature thus far. In the Galaxy are MWC 656 \citep{casares_be-type_2014}, which is presumed to be a Be + BH binary. A candidate system, also in the Galaxy, is HD 130298 \citep{mahy_identifying_2022}, for which the proposed BH companion has a minimum mass of 7 $M_{\odot}$. As for extragalactic examples, an unambiguous case of a quiet O+BH system is VFTS 243 \citep{shenar_x-ray-quiet_2022} in the LMC, along with additional OB+BH candidates VFTS 514 and 779 identified in the same study. There have been several other reported quiet OB + BH systems over the last couple of years, but have been consequently challenged: namely LB-1 \citep{liu_wide_2019,abdul-masih_signature_2020,shenar_hidden_2020}, HR 6819 \citep{rivinius_naked-eye_2020,bodensteiner_is_2020,gies_h_2020,frost_hr_2022}, NGC 1850 BH1 \citep{saracino_black_2022,el-badry_ngc_2022}, and NGC 2004 115 \citep{lennon_vlt-flames_2021,el-badry_ngc_2022} - illustrating the difficulty to find OB + BH systems with certainty. \par 

Given the apparent theoretical frequency of dormant BHs in binary systems, searching for them in single-lined binaries (SB1s) identified in large spectroscopic campaigns of populations of massive stars is a promising method \citep[eg.][]{guseinov_collapsed_1966}. Paper I contains such a sample of SB1s in NGC 6231. An investigation into the potential number of BHs in this cluster has been performed by \cite{van_der_auwera_searching_2017}, which was based on identifying BH candidates through signatures of Bondi-Hoyle accretion (the Carina nebula was also a target of this study). Bondi-Hoyle accretion is the process of isolated BHs accreting from the interstellar medium, so this study probed the single star formation channel for forming stellar mass BHs and did not take into account the various binary channels of BH formation.  First, the current number of isolated BHs in NGC 6231 was analytically estimated through population synthesis computations based on the Kroupa initial mass function \citep[IMF, ][]{kroupa_variation_2001}, and due to the uncertainty on the input parameters (such as cluster age and the slope of the IMF, amongst others), the current number of isolated BHs was estimated to be between 2 -- 8. Then candidates for Bondi-Hoyle accreting BHs were to be identified through the analysis of deep sky X-ray observations made with \textit{XMM-Newton} \citep{sana_xmm-newton_2006}. Through simulating the BH X-ray flux distribution of NGC 6231, it was estimated that none of the Bondi-Hoyle accreting BHs in the cluster would be detected with the XMM observations.   

In this work, we use medium-resolution time-series optical spectroscopy, ground-based and high-precision space based photometry, and state-of-the-art spectral disentangling to characterise the nature of the unseen companions to the SB1 B stars in NGC 6231, and identify any candidates that may harbour compact object companions. The outline of the paper is as follows: Sect. \ref{s:data} describes the target selection, the observations, and the data reduction, Sect. \ref{s:methodology} outlines the methodology used to characterise the unseen companions and the subsequent results, Sect. \ref{s:discussion} contains the discussion of the results and Sect. \ref{s:summary} summarises our findings. \par

%__________________________________________________________________

\section{Observations and data reduction}
\label{s:data}
\subsection{Target selection}
\label{ss:targetselection}

We have selected all of the targets in Paper~I that were reported as SB1s. These 15 targets have radial velocity (RV) measurements and constrained orbital solutions (see Table \ref{table:SB1}), and have all been spectrally classified as having B-type primary stars. Concerning the eccentricities of the orbital solutions, there are several that do not pass the Lucy-Sweeney test \citep{lucy_spectroscopic_1971} at the 5\% significance level. This means that the constrained eccentricity value is not necessary in describing the orbital solution. The 15 targets have also been verified as members of NGC 6231 in Paper~I, using distances estimated from parallaxes in Gaia eDR3 \citep{bailer-jones_estimating_2021}. None of these targets are Be stars, which is notable as the progenitors of B-type star and compact object binary systems are likely to be Be stars. This could be due to the young age of the cluster, as Be stars have been proposed to be (at least partially) the result of binary interaction \citep{bodensteiner_young_2021}.

 \par

\subsection{Observations and data reduction}
\label{sec:observations}
\subsubsection{Spectroscopy}
\label{sss:spectroscopy}

For both the spectral disentangling and the spectroscopic fitting of our targets, we use the same observations that were used in Paper~I. These were carried out with the ESO VLT-FLAMES instrument at the Paranal Observatory, Chile. The instrument was used in its GIRAFFE/MEDUSA + UVES mode, with the purpose of retrieving intermediate resolution ($\lambda / \delta \lambda \sim 6500$) multi-object spectroscopy with the L427.2 (LR02) setting. This gives continuous wavelength coverage between 3964–4567\AA. A total of 31 observations were obtained between 2017 April 02 and 2018 June 17, providing a timebase of 441 days throughout ESO periods 99 and 101\footnote{PI: H. Sana, IDs 099.D-0895 and 0101.D-0163}. The data were reduced following the procedure described in Paper~I, and we refer to it for more details about the observations.

\subsubsection{Photometry}
\label{sss:photometry}

Along with the obtained spectra, we collected photometry that has been previously gathered for the SB1 targets in order to fit each star's spectral energy distribution (SED). We use three catalogues for this purpose. The first is the Two Micron All Sky Survey \citep[2MASS,][]{skrutskie_two_2006}, where observations from two 1.3m telescopes at Mount Hopkins, Arizona and Cerro Tololo, Chile were taken between 1997 June and 2001 February. Photometry is available for each target in near-infrared $J$ (1.25~$\mu$m), $H$ (1.65~$\mu$m) and $K_{s}$ (2.16~$\mu$m) bandpasses. We also use photometry obtained by \citet{sung_initial_2013} with the 1m telescope at Siding Spring Observatory, where photometry was performed on 2000 June 22 and 25. The bandpasses we use in SED fitting are $U$ (0.36~$\mu$m), $B$ (0.44~$\mu$m), $V$ (0.55~$\mu$m) and $I$ (0.80~$\mu$m). The final catalogue that we utilise is Gaia eDR3 \citep{collaboration_gaia_2016,collaboration_gaia_2021}, from which we use photometry recorded in the $G$ (0.62~$\mu$m), $GBp$ (0.51~$\mu$m) and $GRp$ (0.78~$\mu$m) bandpasses. The chosen photometric catalogues provide us ample coverage for the SEDs of the target stars, from ultraviolet (UV) to near-infrared (NIR). A list of magnitudes and the associated uncertainties for each star can be seen in Table \ref{tab:photometry}. \par

\begin{table*}[t]
\centering
\caption{Photometry collected of the targets for SED fitting from \cite{sung_initial_2013}, Gaia eDR3 \citep{collaboration_gaia_2016,collaboration_gaia_2021}, and 2MASS \citep{skrutskie_two_2006}}.\
\label{tab:photometry}
\resizebox{\textwidth}{!}{%
\begin{tabular}{lllllllllll}
\hline\hline
ID                    & U                  & B                  & V                  & I                  & G                  & Gbp                & Grp              & J                & H                & Ks               \\
                      & mag                & mag                & mag                & mag                & mag                & mag                & mag              & mag              & mag              & mag              \\
\hline
NGC 6231 723          & 11.643$\pm$0.010   & 11.896$\pm$0.009   & 11.435$\pm$0.005   & 10.736$\pm$0.009   & 11.325$\pm$0.005   & 11.5506$\pm$0.0029 & 10.838$\pm$0.004 & 10.338$\pm$0.030 & 10.185$\pm$0.035 & 10.092$\pm$0.027 \\
HD 326328             & 9.828$\pm$0.008    & 10.417$\pm$0.006   & 10.213$\pm$0.004   & 9.854$\pm$0.008    & 10.1421$\pm$0.0028 & 10.2483$\pm$0.0029 & 9.907$\pm$0.004  & 9.707$\pm$0.026  & 9.597$\pm$0.021  & 9.549$\pm$0.020  \\
HD 152200             & 7.760$\pm$0.025    & 8.527$\pm$0.007    & 8.391$\pm$0.006    & 8.107$\pm$0.015    & 8.3401$\pm$0.0028  & 8.3999$\pm$0.0030  & 8.171$\pm$0.004  & 8.120$\pm$0.027  & 8.08$\pm$0.07    & 8.06$\pm$0.05    \\
NGC 6231 255          & 12.9950$\pm$0.0014 & 13.1710$\pm$0.0010 & 12.842$\pm$0       & 12.3340$\pm$0.0030 & 12.7292$\pm$0.0028 & 12.9092$\pm$0.0029 & 12.408$\pm$0.004 & 12.045$\pm$0.026 & 11.919$\pm$0.030 & 11.947$\pm$0.027 \\
V* V946 Sco           & 9.833$\pm$0.008    & 10.463$\pm$0.005   & 10.2480$\pm$0.0030 & 9.858$\pm$0.008    & 10.2012$\pm$0.0028 & 10.313$\pm$0.004   & 9.963$\pm$0.004  & 9.728$\pm$0.030  & 9.60$\pm$0.05    & 9.535$\pm$0.020  \\
CD-41 11030           & 8.943$\pm$0.006    & 9.602$\pm$0.005    & 9.4570$\pm$0.0030  & 9.182$\pm$0.008    & 9.4254$\pm$0.0028  & 9.4955$\pm$0.0029  & 9.247$\pm$0.004  & 9.108$\pm$0.023  & 9.064$\pm$0.021  & 9.079$\pm$0.020  \\
NGC 6231 273          & 12.886$\pm$0.012   & 13.063$\pm$0.011   & 12.770$\pm$0.008   & 12.290$\pm$0.012   & 12.6861$\pm$0.0028 & 12.8536$\pm$0.0029 & 12.384$\pm$0.004 & 12.040$\pm$0.030 & 11.925$\pm$0.034 & 11.835$\pm$0.030 \\
CD-41 11038           & 8.154$\pm$0.006    & 8.909$\pm$0.005    & 8.7620$\pm$0.0030  & 8.454$\pm$0.009    & 8.7230$\pm$0.0028  & 8.7961$\pm$0.0029  & 8.530$\pm$0.004  & 8.385$\pm$0.027  & 8.40$\pm$0.04    & 8.361$\pm$0.026  \\
NGC 6231 78           & 11.720$\pm$0.017   & 11.972$\pm$0.015   & 11.635$\pm$0.009   & 11.052$\pm$0.013   & 11.4913$\pm$0.0028 & 11.6884$\pm$0.0029 & 11.137$\pm$0.004 & 10.674$\pm$0.026 & 10.579$\pm$0.027 & 10.501$\pm$0.023 \\
NGC 6231 225          & 13.404$\pm$0.011   & 13.424$\pm$0.009   & 13.0620$\pm$0.0010 & 12.4760$\pm$0.0022 & 12.9700$\pm$0.0028 & 13.177$\pm$0.004   & 12.601$\pm$0.005 & 12.107$\pm$0.029 & 11.92$\pm$0.04   & 11.771$\pm$0.031 \\
V* V1208 Sco          & 9.295$\pm$0.024    & 9.918$\pm$0.023    & 9.708$\pm$0.015    & 9.325$\pm$0.028    & 9.6522$\pm$0.0028  & 9.754$\pm$0.005    & 9.357$\pm$0.005  & 9.009$\pm$0.023  & 9.076$\pm$0.025  & 8.986$\pm$0.021  \\
CXOU J165421.3-415536 & 11.089$\pm$0.028   & 11.497$\pm$0.025   & 11.178$\pm$0.017   & 10.656$\pm$0.029   & 11.0682$\pm$0.0028 & 11.2539$\pm$0.0029 & 10.725$\pm$0.004 & 10.352$\pm$0.023 & 10.234$\pm$0.022 & 10.176$\pm$0.019 \\
CPD-41 7717           & 9.787$\pm$0.007    & 10.404$\pm$0.005   & 10.2030$\pm$0.0030 & 9.873$\pm$0.009    & 10.1494$\pm$0.0028 & 10.2471$\pm$0.0029 & 9.934$\pm$0.004  & 9.696$\pm$0.022  & 9.690$\pm$0.022  & 9.647$\pm$0.020  \\
CPD-41 7722           & 9.601$\pm$0.009    & 10.210$\pm$0.006   & 10.017$\pm$0.004   & 9.707$\pm$0.010    & 9.9559$\pm$0.0028  & 10.0451$\pm$0.0028 & 9.749$\pm$0.004  & 9.568$\pm$0.022  & 9.561$\pm$0.026  & 9.518$\pm$0.025  \\
CPD-41 7746           & 8.755$\pm$0.027    & 9.406$\pm$0.024    & 9.192$\pm$0.015    & 8.821$\pm$0.027    & 9.1280$\pm$0.0028  & 9.2427$\pm$0.0030  & 8.872$\pm$0.004  & 8.646$\pm$0.023  & 8.59$\pm$0.06    & 8.538$\pm$0.019  \\
\hline
\end{tabular}%
}
\end{table*}

We also complement the spectroscopy with space-based photometry, specifically, light curves obtained with the Transiting Exoplanet Survey Satellite \citep[TESS,][]{ricker_transiting_2015}. We use these to further search for non-degenerate companions in our SB1 sample. The 2-min cadence light curves (LCs) were retrieved from the Mikulski Archive for Space Telescopes (MAST3) archive. The light curves are those in the pre-conditioned form (PDCSAP, Pre-search Data Conditioning Simple Aperture Photometry). The 30-min cadence light curves were extracted from the full-frame images (FFIs). Aperture photometry was performed on image cutouts of $50 \times 50$ pixels using the \textsc{PYTHON} package \textsc{LIGHTKURVE}. The source mask was defined from pixels above a given threshold (generally from 3 to 10 depending on the target). The background mask was defined by pixels with fluxes below the median flux, thereby avoiding contamination by nearby field sources. \par

\section{Methodology}
\label{s:methodology}

%\subsection{Orbital solutions}
%\label{ss:orbitalsolutions}
%{\color{ForestGreen}{\bf{That was already presented in Paper I, I would suggest you to integrate that to Spectral disentangling in one or two sentences and remove this subsection. }}}
%First, orbital solutions for each of the targets are needed to spectrally disentangle a potential signature of a faint companion. Orbital parameters for each of our SB1 systems are provided by \cite{banyard_observed_2021}. In this work, the orbital solutions were constrained with the code \texttt{RaV FIT} 
%\citep{mahy_triple_2018}, which compute the orbital parameters given a sample of radial velocity (RV) measurements, which were also obtained by \cite{banyard_observed_2021} through Gaussian/Lorentzian fitting of absorption lines. The parameter estimates were minimised with the Levenberg-Marquadt algorithm and a single systemic velocity ($\gamma$) for each system was considered. Table \ref{table:SB1} details the constrained orbital solutions for the SB1s, and the their RV curves and associated orbital solutions are plotted in Figure~\ref{fig:rvcurves}.

%/STER/gareth/NGC_6231/pdf_summary_paper2.py
\begin{figure*}
\centering
\includegraphics[width=0.9\textwidth]{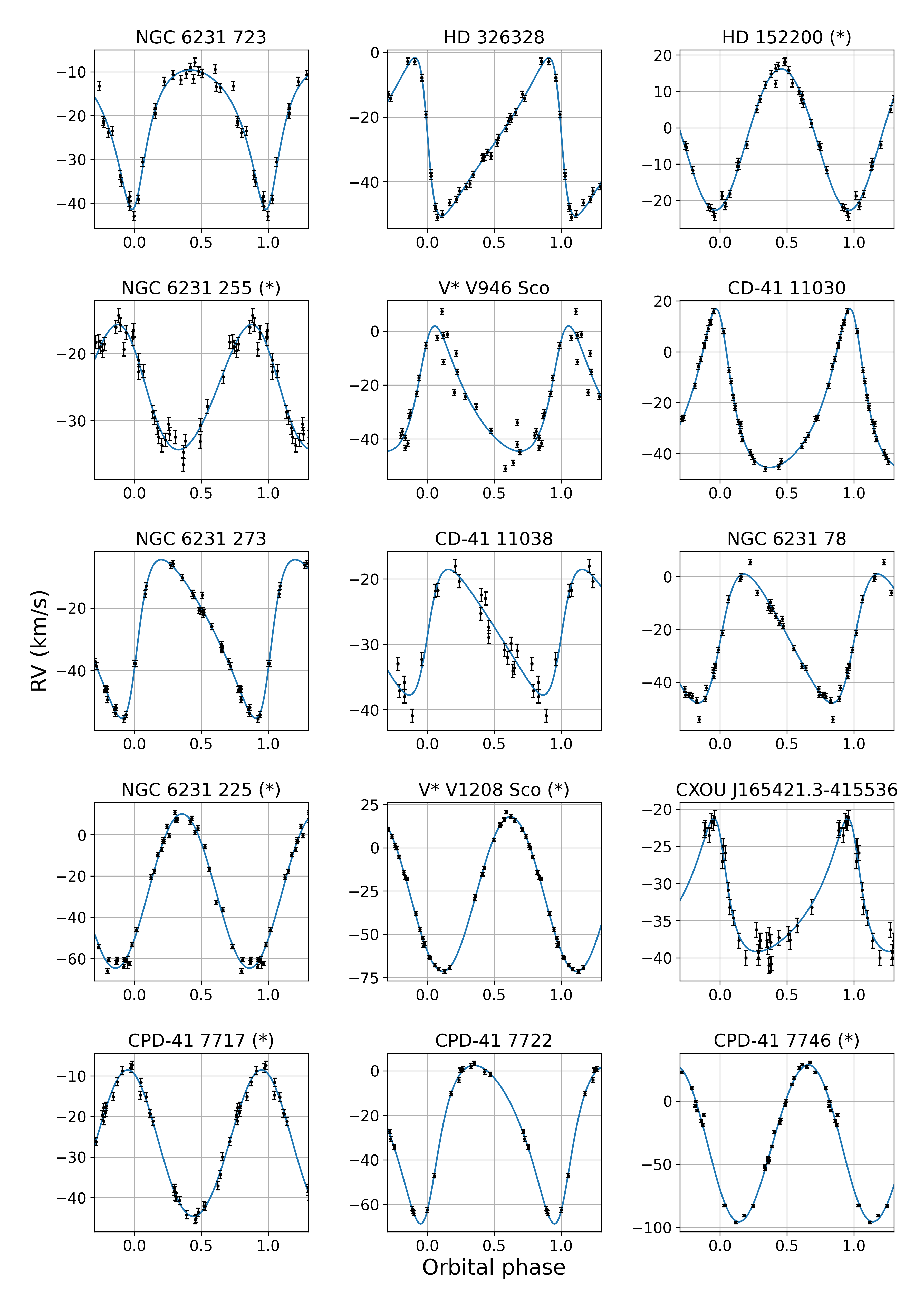}
    \caption{Phase-folded RV curves and orbital solutions constrained by \cite{banyard_observed_2022} for the SB1s. The five marked with (*) do not pass the Lucy-Sweeney test ($e/\sigma_{e}\leq\,2.49$) and so are consistent with being circular.
         }
    \label{fig:rvcurves}
\end{figure*}

\subsection{Spectral disentangling}
\label{ss:disentangling}

To spectrally disentangle a potential signature of a faint companion, orbital solutions for each of the targets are needed. In Paper~I, orbital solutions were constrained by fitting radial velocities (RVs), measured from the multi-epoch FLAMES spectra described in Section~\ref{sss:spectroscopy}, with a damped least squares (DLS) method. Table \ref{table:SB1} details the constrained orbital solutions for the SB1s, and the their RV curves and associated orbital solutions are plotted in Figure~\ref{fig:rvcurves}. With these orbital solutions, we can perform Fourier spectral disentangling \citep{hadrava_orbital_1995} on each target. We use this technique to extract individual spectra of each component from a time series of composite spectra. This way, we can identify signatures of putative faint companions in these systems, and discard the possibility of these systems harbouring compact objects. \par

\begin{table*}[t]
\caption{Orbital solutions for confirmed SB1 systems from \cite{banyard_observed_2022}.}             
\label{table:SB1}      
\centering
\resizebox{\textwidth}{!}{%
%\begin{tabular}{c c c c c l l l l l}
\begin{tabular}{l l l l l l l l l l l l l}
\hline\hline  
Object name           & V (mag) & SpT  & $\chi_2^{red}$ & $T_0 -2450000$              & $P$ (d)               & $e$               & $\omega$ ($^{\circ}$)            & $K_1$ (\kms)     & $\gamma$ (\kms)   & $a_{1}$sin$i$ ($R_{\odot}$) & $f(M) (M_{\odot})$ & RUWE \\
\hline
NGC 6231 723          & 11.41   & B3V   & 3.4            & 7896.4$\pm$0.4   & 30.93$\pm$0.05    & 0.39$\pm$0.03   & 14$\pm$5    & 15.9$\pm$0.5 & -19.5$\pm$0.4   & 9.0$\pm$0.3                 & 0.010$\pm$0.001 & 23.41     \\
HD 326328             & 10.213  & B2V   & 1.2            & 7888.6$\pm$0.1   & 28.417$\pm$0.01   & 0.49$\pm$0.01   & 85$\pm$2    & 24.4$\pm$0.4 & -27.2$\pm$0.2   & 11.9$\pm$0.2                & 0.028$\pm$0.001 & 0.97     \\
HD 152200             & 8.391   & B0V   & 2.2            & 7901.2$\pm$0.4   & 4.4440$\pm$0.0007 & 0.04$\pm$0.02*   & 190$\pm$30  & 19.5$\pm$0.4 & -2.6$\pm$0.3    & 1.71$\pm$0.03               & 0.0034$\pm$0.0002   & 0.86  \\
NGC 6231 255          & 12.842  & B9V   & 2.6            & 7899.2$\pm$0.8   & 9.271$\pm$0.008   & 0.09$\pm$0.05*   & 50$\pm$30   & 9.3$\pm$0.4  & -25.5$\pm$0.3   & 1.71$\pm$0.07               & 0.0008$\pm$0.0001   & 1.05   \\
V* V946 Sco           & 10.281  & B2V   & 32.6           & 7935$\pm$3       & 97.9$\pm$0.8      & 0.32$\pm$0.07   & 140$\pm$10  & 23$\pm$2     & -27$\pm$1       & 43$\pm$4                    & 0.11$\pm$0.03   & 0.93       \\
CD-41 11030           & 9.457   & B0.5V & 0.8            & 7869.2$\pm$0.3   & 70.06$\pm$0.05    & 0.369$\pm$0.008 & 24$\pm$1    & 31.2$\pm$0.3 & -24.8$\pm$0.2   & 40.1$\pm$0.4                & 0.176$\pm$0.006 & 1.00     \\
NGC 6231 273          & 12.770   & B9V   & 1.9            & 7866.45$\pm$0.09 & 13.948$\pm$0.005  & 0.36$\pm$0.02   & 67$\pm$3    & 25.4$\pm$0.5 & -26.4$\pm$0.3   & 6.5$\pm$0.2                 & 0.019$\pm$0.001 & 0.91     \\
CD-41 11038           & 8.756   & B0V   & 4.5            & 7881$\pm$4       & 83.9$\pm$0.9      & 0.33$\pm$0.07   & 80$\pm$10   & 9.6$\pm$0.8  & -27.9$\pm$0.5   & 15$\pm$1                    & 0.006$\pm$0.002 & 0.86     \\
NGC 6231 78           & 11.585  & B4V   & 5.0            & 7896.2$\pm$0.4   & 31.07$\pm$0.03    & 0.25$\pm$0.03   & 87$\pm$6    & 24.4$\pm$0.7 & -23.1$\pm$0.4   & 14.5$\pm$0.4                & 0.042$\pm$0.004 & 0.81     \\
NGC 6231 225          & 13.062  & B9V   & 8.2            & 7901$\pm$4       & 7.768$\pm$0.002   & 0.01$\pm$0.02*   & 200$\pm$200 & 37.4$\pm$0.7 & -27.1$\pm$0.5   & 5.7$\pm$0.1                 & 0.042$\pm$0.003 & 1.04     \\
V* V1208 Sco          & 9.708   & B0.5V & 1.7            & 7900.1$\pm$0.7   & 5.2193$\pm$0.0003 & 0.010$\pm$0.008* & 140$\pm$50  & 44.7$\pm$0.4 & -26.5$\pm$0.3   & 4.61$\pm$0.04               & 0.048$\pm$0.001 & 0.82     \\
CXOU J165421.3-415536 & 11.178  & B2V   & 2.1            & 7943$\pm$4       & 204$\pm$7         & 0.41$\pm$0.05   & 40$\pm$10   & 9.0$\pm$0.4  & -33.0$\pm$0.4   & 33$\pm$2                    & 0.012$\pm$0.002 & 0.95     \\
CPD-41 7717           & 10.21   & B2V   & 1.4            & 7898.4$\pm$0.2   & 2.5080$\pm$0.0001 & 0.04$\pm$0.02*   & 200$\pm$20  & 18.0$\pm$0.3 & -27.3$\pm$0.2   & 0.89$\pm$0.02               & 0.00152$\pm$0.00008 & 0.94 \\
CPD-41 7722           & 10.017  & B0V   & 1.1            & 7894.4$\pm$0.2   & 27.877$\pm$0.009  & 0.29$\pm$0.01   & 32$\pm$2    & 35.5$\pm$0.4 & -24.5$\pm$0.3   & 18.7$\pm$0.2                & 0.114$\pm$0.004 & 0.83     \\
CPD-41 7746           & 9.24    & B0.2V & 11.8           & 7897.2$\pm$0.6   & 6.3498$\pm$0.0008 & 0.03$\pm$0.01*   & 130$\pm$30  & 62$\pm$1     & -32.3$\pm$0.7   & 7.8$\pm$0.1                 & 0.157$\pm$0.008 & 0.94    \\
\hline
\end{tabular}%
}
\flushleft
\footnotesize * Eccentricities that are not statistically significant as they do not pass the Lucy-Sweeney test ($e/\sigma_{e}\leq\,2.49$)
\end{table*}

For this purpose, we apply the Fourier spectral disentangling in a similar manner as to what was done by Mahy et al. (2022, in press) and Shenar et al. (2022, in press), where the method was used to characterise unseen companions in Galactic and LMC O-type SB1s. The orbital parameters are used as input into finding a self-consistent solution from a time-series of composite spectra obtained for each target, and this optimisation is done with a Nelder \& Mead simplex \citep{nelder_simplex_1965} to find the best solution in a multidimensional space. \par

There will be a lower limit on the mass of an extracted companion in this disentangling process, and we will be unable to extract the spectral signatures of companions fainter than a certain fraction of the brightness of the primary. This should, by the nature of the primary masses in our sample, be lower than for the systems in the Mahy et al. (in prep) study, where the lower limit of unseen companions extracted is about 4-5 $M_{\odot}$.\par 

After applying spectral disentangling, we now have three categories in which we can place the targets that were identified as SB1s in Paper~I. We have those that were found to have a signature of a secondary companion, and as such are newly-identified SB2s. These are seven of the 15 targets - NGC 6231 723, HD 326328, CD$-$41 11030, NGC 6231 78, V*V1208 Sco, CPD$-$41 7722, and CPD$-$41 7746. As we identified lower-mass stellar companions in these systems, we discount the possibility of them harbouring a compact object as a companion. \par 

For the remaining eight objects, we tried further to isolate the spectroscopic signature of a putative companion. Since the multidimensional disentangling fit fails to recover the secondary signal, we fixed the orbital period ($P$), eccentricity ($e$), longitude of the periastron passage ($\omega$), the time of reference ($T_{0}$), and the RV semi-amplitude $K_{1}$ of the primary star, only allowing to $K_{2}$ to vary. This grid-based disentangling approach \citep{fabry_resolving_2021} is similar to that successfully applied to detect spectral signatures of faint companions with optical brightness contributions below 1\% \citep{shenar_hidden_2020,bodensteiner_is_2020}. While we were able to extract the spectra of the putative companions and characterise their main features, retrieving their semi-amplitude $K_2$ with accuracy was more challenging - typical uncertainties on the values of K2 tended to be around 50 \kms. This, however, does not impact extracting the spectrum of the companions.

A second category of targets are those that we classified as ambiguous cases, meaning that a secondary object was extracted from the composite spectra, but further attention or data is required to allow their characterisation. For example. the extracted spectra of a secondary companion may not resemble a stellar spectrum, or we might detect Balmer lines in the spectrum without being able to characterise the companion. This could be due to the potential companion being too faint for the limits of this methodology. From the second category, there are 4 targets - V* V946 Sco, CD-41 11038, CXOU J165421.3-415536, and NGC 6231 225. Finally, the third category are targets where no companion could be identified through spectral disentangling, of which there are 4 - HD 152200, CPD-41 7717, NGC 6231 255, and NGC 6231 271. In these cases, the output spectra appear featureless. \par

As a result, we subject targets in these latter two categories (both the ambiguous cases, and those which we have not been able to identify as SB2s) to more rigorous study to see whether the possibility of a compact object companion can be discounted. 

We performed extensive simulations to determine the sensitivity of the spectral disentangling. The goal of these simulations is to be able to exclude the presence of a non-degenerate companion down to certain mass and certain flux ratios. For our simulations, we started with considering binary systems where the properties are similar to those of our visible stars (see Table~\ref{tab:sb1stellarparams}). Given that the masses of the visible stars are just an indication to compute the mass ratio with the mock object, we consider the average estimate between the spectroscopic and the evolutionary masses. To build the mock composite spectra, we use the observed composite spectra and we inject a synthetic spectrum (with a matched S/N ratio to the observations) with the characteristics of the mock objects, scaled by the flux ratios, and shifted by the appropriate radial velocities. 
To estimate the properties of the mock stars (spectral types, effective temperatures, masses, and absolute visual magnitudes), we use the tables from \citet{schmidt-kaler_automated_1982}. The stars listed in  Table~\ref{tab:sb1stellarparams} are all classified as B0V or B2V. The absolute $V$-band magnitude of a B0V star is predicted to $M_V = -4.0$, while that of a B2V star is $M_V = -2.45$. \par 

\begin{figure}
\centering
\includegraphics[width=\hsize]{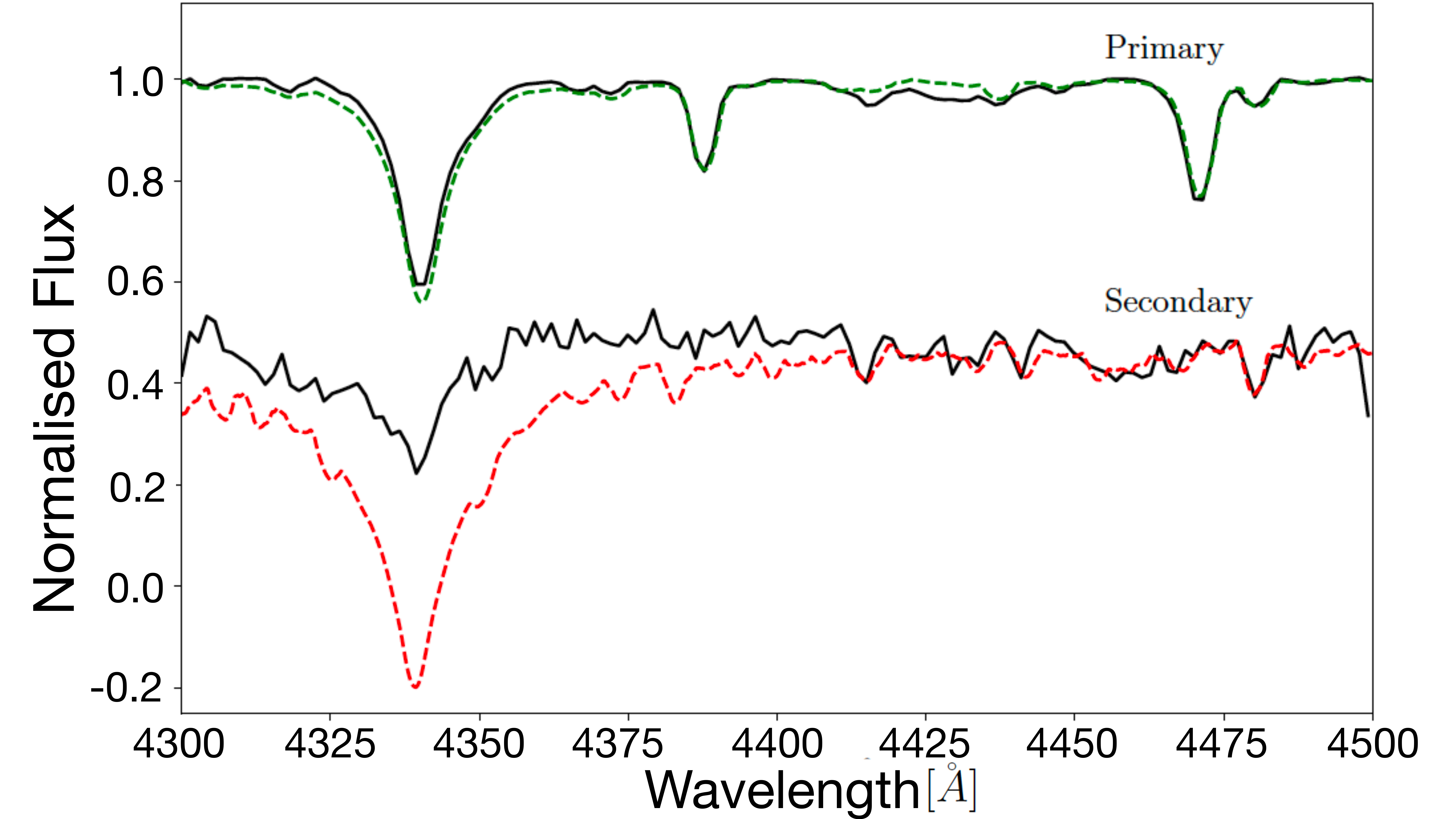}
    \caption{Results from simulated spectra of a mock B2V $+$ A7V binary mimicking the real orbit and data of CPD~$-41~7717$. In green and in red, the synthetic spectra (templates) computed from the stellar parameters of the primary and the secondary components, respectively. The spectra of the secondary have been shifted vertically for clarity. }
    \label{fig:simulationB2A7}
\end{figure}

For the B0V primaries, we were able to extract the spectrum of the secondary object down to a spectral type B9V or A0V, i.e., objects with flux ratios of about 1\%--1.5\% and mass ratios of about 0.10--0.15. This corresponds to secondaries with absolute $V$ magnitude of $M_V = 0.2 - 0.65$ and masses of roughly 2.3--2.5~$M_{\odot}$. When the primaries are classified as B2V, we managed to extract the mock spectra of the secondaries down to spectral types A5V or A7V, i.e., objects with mass and flux ratios of about 0.02 and 0.015, respectively. This corresponds to absolute $V$ magnitude of $M_V = 1.95 - 2.2$ and masses between 1.4 and 2.0~$M_{\odot}$ (Fig.~\ref{fig:simulationB2A7}). \par

We also consider the possibility that the non-detected companions could be stripped stars. We adopted the parameters given by and models produced by \citet{gotberg_spectral_2018} at solar metallicity. When the visible primary star is classified as B0V, our observed data would be of sufficient quality to extract a stripped star companion with a temperature $T_{\rm eff} = 62$~kK and a radius of $R \sim 0.55~R_{\odot}$, i.e., an object with an initial mass of about $M_{\rm ini} \sim 8.5~M_{\odot}$ and a stripped mass of $M_{\rm strip} \sim 2~M_{\odot}$. When the visible primary star is classified as B2V, we can extract the spectrum of a stripped star down to an $T_{\rm eff} = 43$~kK and a radius of $R \sim 0.34~R_{\odot}$, i.e., an object with an initial mass of about $M_{\rm ini} \sim 4.5~M_{\odot}$ and a stripped mass of about $M_{\rm strip} \sim 1~M_{\odot}$. It should be noted that stripped stars of $M_i = 4.5 M_{\odot}$ form the earliest after 100 Myr, and so NGC 6231 should not contain such objects. \par

Another possibility is that the B star component can be spun up due to mass transfer from a stripped star component. In this case, due to the rotational broadening of the spectral component of the B star, it would be the stripped star that would dominate the composite spectrum. However, as none of the spectra resemble stripped star spectra, we can discard this.

The last case that we considered for the is the case of a triple system, with two tight low-mass stars forming the inner system and the B-type stars orbiting around with a wider orbital period. While it is difficult for HD~152200 and CPD~$-41~7717$ to be considered as triple systems given their short orbital period, the three other objects listed in Table~\ref{tab:sb1stellarparams} have all orbital period longer than 80 days. According to \citet{toonen_evolution_2020} \citep[based on the stability criterion described by][]{mardling_dynamics_1999,mardling_tidal_2001}, a triple system, composed of an inner system with two A7V stars or later and an outer B-type star orbiting with the parameters given in Table~\ref{table:SB1}, would be considered as stable. If the two inner stars are earlier than A7V, the triple system would be unstable. We also consider the light ratios involved in such systems, where the two stars in the inner system have a similar light contribution, and then we compare with the B star in the outer system. Each low-mass star would contribute between 0.3 and 1.2\% to the total flux (depending whether we consider a B0V primary or a B2V). Therefore, F stars in an inner close system would be extremely difficult to detect, but late A-type stars would be detectable. However, the resolution of our composite spectra would not be sufficient to extract the individual contribution of the low-mass stars, but rather a merged spectrum. \par

\subsection{Atmosphere modelling}
\label{ss:atmospheremodelling}

To constrain the mass of the unseen secondary using the binary mass function (Section~\ref{ss:minimummasses}), we first need an estimate of the primary mass. We choose to do this in two ways - by estimating the spectroscopic mass of the primary star, and by estimating the evolutionary mass of the primary star. The spectroscopic mass is named as such as it is computed with stellar gravities and radii ($M_{\mathrm{spec}} \propto gR^{2}$) determined through the fitting of stellar spectra and SEDs. The evolutionary mass, on the other hand, is inferred by comparing the target's properties, namely log($L/L_{\odot}$), $T_{\mathrm{eff}}$, log $g$, and $v\,\sin\,i$, to stellar evolutionary tracks. \par 

For both of these estimates, we first need to estimate the stellar parameters of the primary star. Specifically, we want to obtain the effective temperature ($T_{\mathrm{eff}}$), the surface gravity ($g$), the projected rotational velocity ($v\,\sin\,i$), the stellar radius ($R_{*}$), and the bolometric luminosity ($L_{bol}$) of the primary star, along with the extinction to the star, expressed in this Section~as the colour excess, $E(B-V)$ We use cgs units unless indicated otherwise. \par 

We can obtain estimates to these parameters by adopting a grid-oriented fitting approach, based on synthetic TLUSTY \citep{hubeny_non-lte_1995} models that relax the assumption of local thermodynamic equilibrium (non-LTE or NLTE), and following the implementation developed by Bodensteiner et al. (in prep). \par 

\subsubsection{Spectroscopic fitting}
\label{sss:spectro_fit}

In this implementation, we first combined all the observed composite spectra of an SB1 after having shifted them with their RVs (measured in Paper I) to put them in the same reference frame, and then we have co-added them to produce a master spectrum of the primary star with a higher signal-to-noise ratio (S/N) than in the individual epochs. We then fitted these master spectra with a TLUSTY model degraded to the resolution of our observed spectra ($\lambda / \delta \lambda \sim 6500$) and rebinned to the wavelength step of our data. We use both the BSTAR2006 \citep{lanz_grid_2007} and OSTAR2002 \citep{lanz_grid_2003} grids to include the hottest stars in our sample. The BSTAR2006 grid contains synthetic spectra for effective temperatures from 15000 to 30000 K in steps of 1000 K, and log $g$ from 1.75 to 4.75 with 0.25 dex steps. The OSTAR2002 grid's synthetic spectra have effective temperatures ranging from 27500 K to 55000 K with steps of 2500 K (we truncate the grid here to 30000 K as effective temperatures below this are already present in the BSTAR2006 grid with a smaller step), and log $g$ from 3.0 to 4.75 in 0.25 dex steps. Both grids have solar metallicities. The BSTAR2006 grid spectra have a microturbulence of 2 \kms, whereas those in OSTAR2002 have a microturbulence of 10 \kms. Spectra in both grids have a coverage of the entire UV to optical range (900 to 10000 \AA). \par 

The spectroscopic fitting allows us to constrain $T_{\mathrm{eff}}$, log $g$, and $v\,\sin\,i$ from each target's master spectrum. To account for $v\,\sin\,i$, we pre-broaden each TLUSTY model with a range of $v\,\sin\,i$ values from 20 to 400 \kms. The best-fitting model is then constrained by locating the minimum of the $\chi^2$-distribution for each of the three parameters. 1-$\sigma$ errors are calculated from the estimated 68.2 \% confidence interval on the parameter range. \par 

We run into the limitations of using the TLUSTY grid when we consider the targets in our sample that have effective temperatures below 15000 K (i.e. are later than spectral type B5), of which there are three (NGC 6231 255, NGC 6231 273, and NGC 6231 225, all classified as B9 stars). For these stars, GSSP \citep{tkachenko_grid_2015} or other databases of synthetic models matching low-mass star properties such as, for example, \cite{kurucz_model_1979}, ATLAS9 \citep{castelli_new_2003}, or PHOENIX \citep{allard_phoenix_2016}, should be used for constraining their stellar parameters. However, based on the expected mass of a B9 star and the very small binary mass functions of these SB1s (Table \ref{table:SB1}), we do not consider these as likely candidates for harbouring compact objects, and so we exclude them from further study in this work. This reduces the sample of potential candidate systems in this work to five. However, finding such a low-mass BH binary would be a potential progenitor of known low-mass BH X-ray binaries, and so such a find would be interesting.  \par

\subsubsection{Photometric fitting}
\label{sss:combinedfit}

Alongside the spectroscopic fitting, like in the original implementation by Bodensteiner et al. (in prep), we also fit the synthetic spectral energy distributions (SEDs) provided for each TLUSTY grid to observed photometry of various catalogues (Table~\ref{tab:photometry}). The photometry we used is chosen to cover the ultra-violet (UV) to the near-infrared (NIR) domain of the SED of each B star. As described in Section~\ref{sec:observations}, the magnitudes and their associated uncertainties in the $J$, $H$, and $K_{s}$ bands are retrieved from 2MASS \citep{skrutskie_two_2006}, those in the $U$, $V$, $B$ and $I$ bands were obtained by \cite{sung_initial_2013}, and $G$, $GBp$, $GRp$ magnitudes were retrieved from Gaia eDR3 \citep{collaboration_gaia_2016,collaboration_gaia_2021}.  \par 

We use the same grid-search approach as the spectroscopic fit, where our best-fitting model is constrained through locating the minimum of the $\chi^2$-distribution of several parameters. The same parameter spaces are explored as described in Sect.~\ref{sss:spectro_fit} for $T_{\mathrm{eff}}$ and log $g$. The constraints on $T_{\mathrm{eff}}$ and (especially) log $g$ are weaker here than for the spectroscopic fits, but the advantage of using the photometric fit is in constraining the stellar radius, and therefore the luminosity, of the target. For the photometric fit, we scale the TLUSTY model grid to stellar radii from 3 to 12 $R_{\odot}$ in steps of 0.1 $R_{\odot}$, and colour excesses from 0 to 0.9 mag in steps of 0.1 mag. To implement these further two parameter ranges in the grid, for every model in the grid of $T_{\mathrm{eff}}$ and log $g$ we scaled the flux at the stellar surface with the distance to NGC 6231 \citep[1.71 kpc,][]{kuhn_kinematics_2019} and then to each stellar radius in the radius grid. Then, interstellar exctinction is applied to every model through the python package \texttt{DUST EXTINCTION}\footnote{\url{https://dust-extinction.readthedocs.io/en/stable/}} in the color excess range defined above. This gives us a 4-dimensional grid of SED models to fit the observed photometry with observed photometry of various catalogues. \par

\subsubsection{Combined fit}
\label{sss:combinedfit}

To find the best-fitting model for all the parameters explored between the spectroscopic and photometric fitting, and again in a similar manner to Bodensteiner et al. (in prep) we combine the computed $\chi^2$ from each independent fit of each model with common values of $T_{\mathrm{eff}}$ and log $g$, but varying values of $v\,\sin\,i$, radius, and colour excess. The result is a 5-dimensional grid, with the parameters of $T_{\mathrm{eff}}$, log $g$, $v\,\sin\,i$, radius, and colour excess. The constrained stellar parameters for the SB1 candidates identified in Section~\ref{ss:disentangling} are presented in Table \ref{tab:sb1stellarparams}.

\begin{table*}[t]
\centering
\caption{Constrained stellar parameters from combined spectroscopic and photometric fitting (Sec.~\ref{sss:combinedfit}) and the evolutionary fitting (Sec.~\ref{ss:evofitting}) for the SB1s identified as potentially harbouring compact objects (in Section~\ref{ss:disentangling}).}
\label{tab:sb1stellarparams}
\resizebox{\textwidth}{!}{%
\begin{tabular}{lllllllllll}
\hline\hline
ID                    & SpT & $T_{\mathrm{eff}}$               & log $g$                         & $v\,\sin\,i$                & $R_{p}$                       & E(B-V)                        & $M_{spec}$                         & log($L/L_{\odot}$)               & $M_{evo}$                        & Age                             \\
                      &     & K                                & dex                             & \kms{}                      & $R_{\odot}$                       & mag                           & $M_{\odot}$                        & dex                              & $M_{\odot}$                      & Myr                               \\
\hline
HD 152200             & B0V & 30000 $\substack{+1200 \\ -100}$ & 4.0 $\substack{+0.3 \\ -0.3}$ & 240 $\substack{+10 \\ -20}$ & 7.6 $\substack{+1.3 \\ -0.1}$ & 0.4 $\substack{+0.1 \\ -0.1}$ & 23.4 $\substack{+15.6 \\ -13.6}$ & 4.62 $\substack{+0.2 \\ -0.02}$  & 17.8 $\substack{+1.2 \\ -0.8}$ & 5.7 $\substack{+0.4 \\ -1.1}$  \\
V* V946 Sco           & B2V & 23000 $\substack{+1000 \\ -300}$ & 4.00 $\substack{+0.02 \\ -0.20}$ & 220 $\substack{+10 \\ -30}$ & 5.0 $\substack{+0.1 \\ -0.8}$ & 0.5 $\substack{+0.1 \\ -0.1}$ & 10.4 $\substack{+0.6 \\ -4.9}$  & 3.80 $\substack{+0.09 \\ -0.20}$  & 9.4 $\substack{+0.6 \\ -0.4}$  & 11.3 $\substack{+2.4 \\ -2.2}$ \\
CD-41 11038           & B0V & 32500 $\substack{+100 \\ -2000}$ & 4.25 $\substack{+0.05 \\ -0.10}$ & 100 $\substack{+10 \\ -10}$ & 7.0 $\substack{+0.1 \\ -0.1}$ & 0.5 $\substack{+0.1 \\ -0.1}$ & 32.1 $\substack{+3.8 \\ -7.4}$  & 4.69 $\substack{+0.01 \\ -0.10}$ & 17.2 $\substack{+0.9 \\ -0.9}$  & 2.7 $\substack{+1.2 \\ -1.4}$   \\
CXOU J165421.3-415536 & B2V & 20000 $\substack{+1000 \\ -600}$ & 4.00 $\substack{+0.01 \\ -0.30}$ & 160 $\substack{+20 \\ -20}$ & 4.3 $\substack{+0.1 \\ -0.7}$ & 0.6 $\substack{+0.1 \\ -0.2}$ & 7.3 $\substack{+0.4 \\ -5.2}$   & 3.4 $\substack{+0.1 \\ -0.2}$    & 7.0 $\substack{+0.6 \\ -0.4}$  & 19.1 $\substack{+6.6 \\ -9.8}$ \\
CPD-41 7717           & B2V & 21000 $\substack{+600 \\ -300}$  & 3.80 $\substack{+0.02 \\ -0.20}$ & 200 $\substack{+20 \\ -20}$ & 4.7 $\substack{+0.8 \\ -0.1}$ & 0.4 $\substack{+0.1 \\ -0.1}$ & 6.1 $\substack{+1.9 \\ -2.4}$   & 3.59 $\substack{+0.20 \\ -0.04}$ & 8.2 $\substack{+0.6 \\ -0.3}$   & 20.0 $\substack{+2.1 \\ -3.1}$ \\
\hline
\end{tabular}
}
\end{table*}

The spectroscopic mass estimates provided in Table \ref{tab:sb1stellarparams} are calculated from the gravities and radii constrained for each star from the combined fit.

\subsection{Evolutionary fitting}
\label{ss:evofitting}

Now that we have constrained the stellar parameters of the primary stars of our sample, we can provide a second estimate of the primary mass through evolutionary fitting. We use the obtained  log($L/L_{\odot}$), $T_{\mathrm{eff}}$, log $g$, and $v\,\sin\,i$ as input for the code BONNSAI \citep[BONN Stellar Astrophysics Interface,][]{schneider_bonnsai_2014,schneider_bonnsai_2017}, which uses Bayesian statistics to compare observed stellar parameters to evolutionary stellar models, and provides predicted evolutionary parameters for the targets, including the evolutionary mass and the age of the star. It should be noted that these estimates will only be consistent with systems that have not interacted before. In this work, we compare our inputted parameters against the BONN single-star evolutionary models of Galactic metallicity \citep{brott_rotating_2011}. The predicted evolutionary masses and ages are presented in Table \ref{tab:sb1stellarparams}. \par

\begin{figure}
\centering
\includegraphics[width=\hsize]{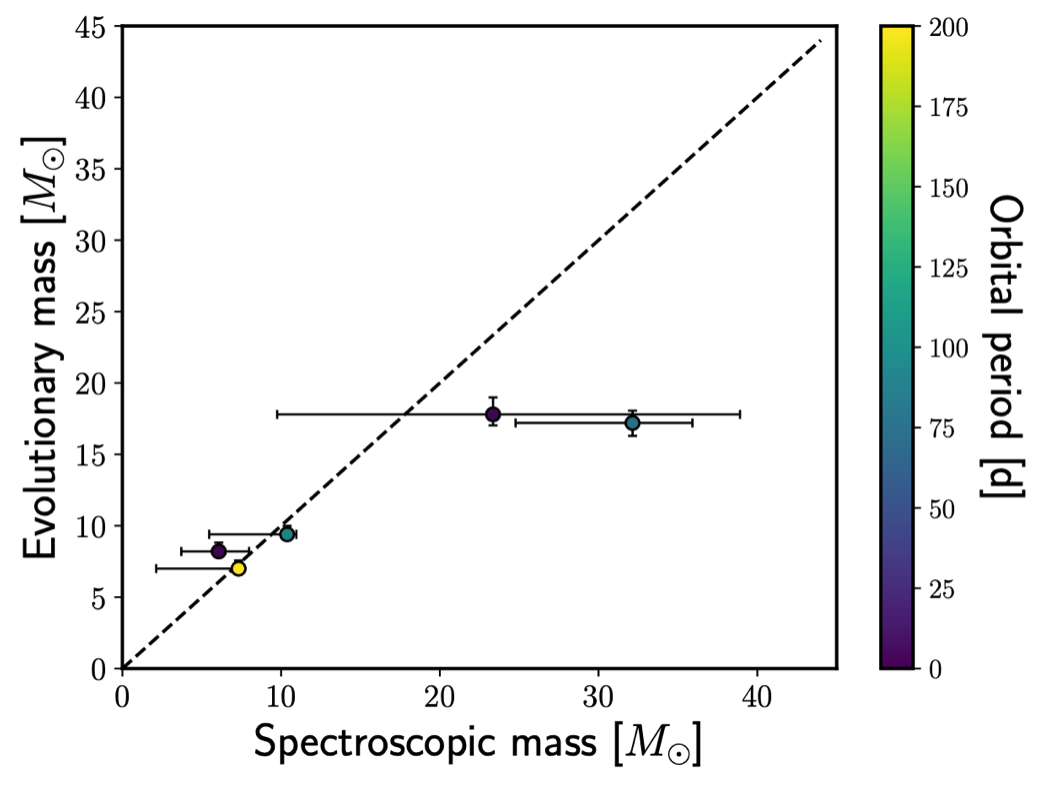}
    \caption{Comparison of the spectroscopic and evolutionary estimates of the compact object candidate SB1s, along with their orbital periods. The dashed line indicates agreement between the two estimates.  
         }
    \label{f:massdiscrepancy}
\end{figure}

There is a potential mass discrepancy present between the spectroscopic and evolutionary estimates of the primary masses of these stars (Figure~\ref{f:massdiscrepancy}) at higher masses, though the large errors on the spectroscopic masses of HD 152200 (and to an extent, CD-41 11038) make this unclear. The discrepancy in the estimation of stellar masses has long been an issue \citep{herrero_intrinsic_1992,burkholder_mass_1997,weidner_masses_2010,mahy_tarantula_2020}. The mass discrepancy found in intermediate- and high-mass eclipsing SB2 stars has been discussed by \cite{tkachenko_mass_2020}, where it was concluded that the discrepancy was a result of an under-estimation of the convective core mass of these stars and the neglect of high micro-turbulent velocities and high turbulent pressure in stellar atmosphere models in the spectral analysis of these stars. However, robustly studying the mass discrepancy in our own targets is out of the scope of this work, and so we consider both estimates of the primary mass going forward. \par

There is also a discrepancy between the estimated age of NGC 6231 and the ages of some of the SB1s. Ages for the cluster have been derived between 2 \citep{sana_xmm-newton_2006,sana_xmm-newton_2007} and 7 \citep{sung_initial_2013} Myr. However, V* V926 Sco, CXOU J165421.3-415536 and CPD-41 7717 all have evolutionary age estimates above this range. This is especially true in the case of CPD-41 7717, which has an estimated evolutionary age of 20.0 $\substack{+15.6 \\ -13.6}$ Myr. Accretion from a donor and consequential rejuvenation effects would result in the star appearing younger than expected, but it becomes more difficult to explain stars that appear older than expected. This could be a signature of more continuous star formation in the cluster.

\subsection{Minimum masses of unseen companions}
\label{ss:minimummasses}

With the estimated primary masses (both spectroscopic and evolutionary) obtained, the ranges of masses of the unseen companions can be calculated in the same manner as \citep{mahy_identifying_2022}. We compute the minimum companion mass for the SB1 systems for which no signature of a companion has been extracted with the spectral disentangling. For that purpose, we use the binary mass function $f(m)$, and the spectroscopic and evolutionary masses (estimated in Sect. \ref{ss:atmospheremodelling} and \ref{ss:evofitting} respectively) of the primary stars:
\begin{equation}
f(m) \equiv \frac{M^{3}_{u}\sin^{3}i}{(M_{u} + M_{p})^2} = \frac{P_{\mathrm{orb}}(1-e^2)^{3/2}K^3}{2{\pi}G}
\label{bmf}
\end{equation}
where $M_{u}$ is the mass of the unseen companion, $M_{p}$ the mass of the primary star, $i$ the inclination, $P_{\mathrm{orb}}$ the orbital period of the binary system, $e$ the eccentricity of the orbit, $K$ the RV semi-amplitude of the primary star, and $G$ the gravitational constant. As $\sin\,i$ lies between 0 and 1, we can write $f(m)$ as an inequality:
\begin{equation}
\frac{M^{3}_{u}\sin^{3}i}{(M_{u} + M_{p})^2} = \frac{P_{\mathrm{orb}}(1-e^2)^{3/2}K^3}{2{\pi}G} \le \frac{M^{3}_{u}}{(M_{u} + M_{p})^2}
\label{bmf2}
\end{equation}
and so we can now solve this inequality to get two estimates of the minimum mass of the unseen companion, one with the spectroscopic estimate for the primary mass and one with the evolutionary estimate of the primary mass. \par 
We can also independently constrain the lower limit of the inclination, $i$ of each system. To do this, we estimate each primary star's critical rotational velocity, $v_{\mathrm{crit}}$, which we define as:
\begin{equation}
v_{\mathrm{crit}} = \sqrt{\frac{2GM_{p}}{3R_{p}}}
\label{vcrit}
\end{equation}
where $G$ is the gravitational constant, $M_{p}$ the mass of the primary star, and $R_{p}$ the radius of the primary star. Without information on the inclinations of the systems in our sample, and with the TLUSTY models we use assuming spherical symmetry of the stars, we make the assumption that the stellar radii, $R_{p}$ that we constrain through our analysis are equal to the polar radii of the stars. We also make the assumption that the rotational axes are perpendicular to the orbital plane, and that the maximum rotational velocity of the star is the computed critical velocity (Equation \ref{vcrit}), then it follows that
\begin{equation}
\frac{v\,\sin\,i}{\sin\,i} \leq v_{\mathrm{crit}}.
\label{inclination}
\end{equation}
From this we can constrain a minimum inclination for each binary system, and consequently a maximum mass of the unseen companion of each system. This assumption does not hold for systems that have NS companions (and potentially those with BH companions) due to the misalignment of the rotational and orbital axes due to a supernova kick. This impacts only the upper limit on the predicted mass of the unseen companion, and so does not impact our conclusions. \par
\begin{figure*}
    \centering
        \begin{subfigure}[b]{0.45\hsize}
            \centering
            \includegraphics[width=\hsize]{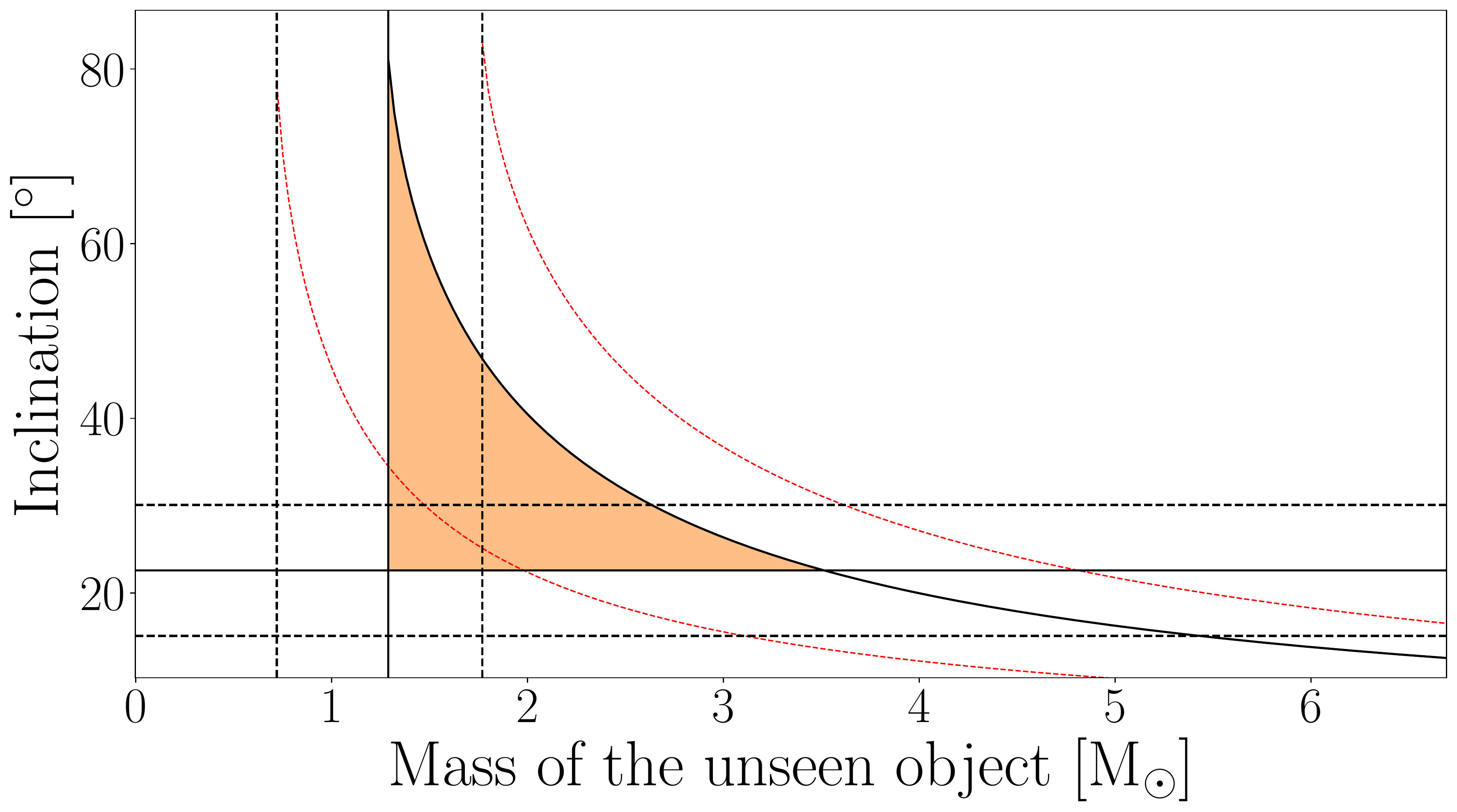}
            \caption{HD 152200}
        \end{subfigure}
        \hfill
        \begin{subfigure}[b]{0.45\hsize}
            \centering
            \includegraphics[width=\hsize]{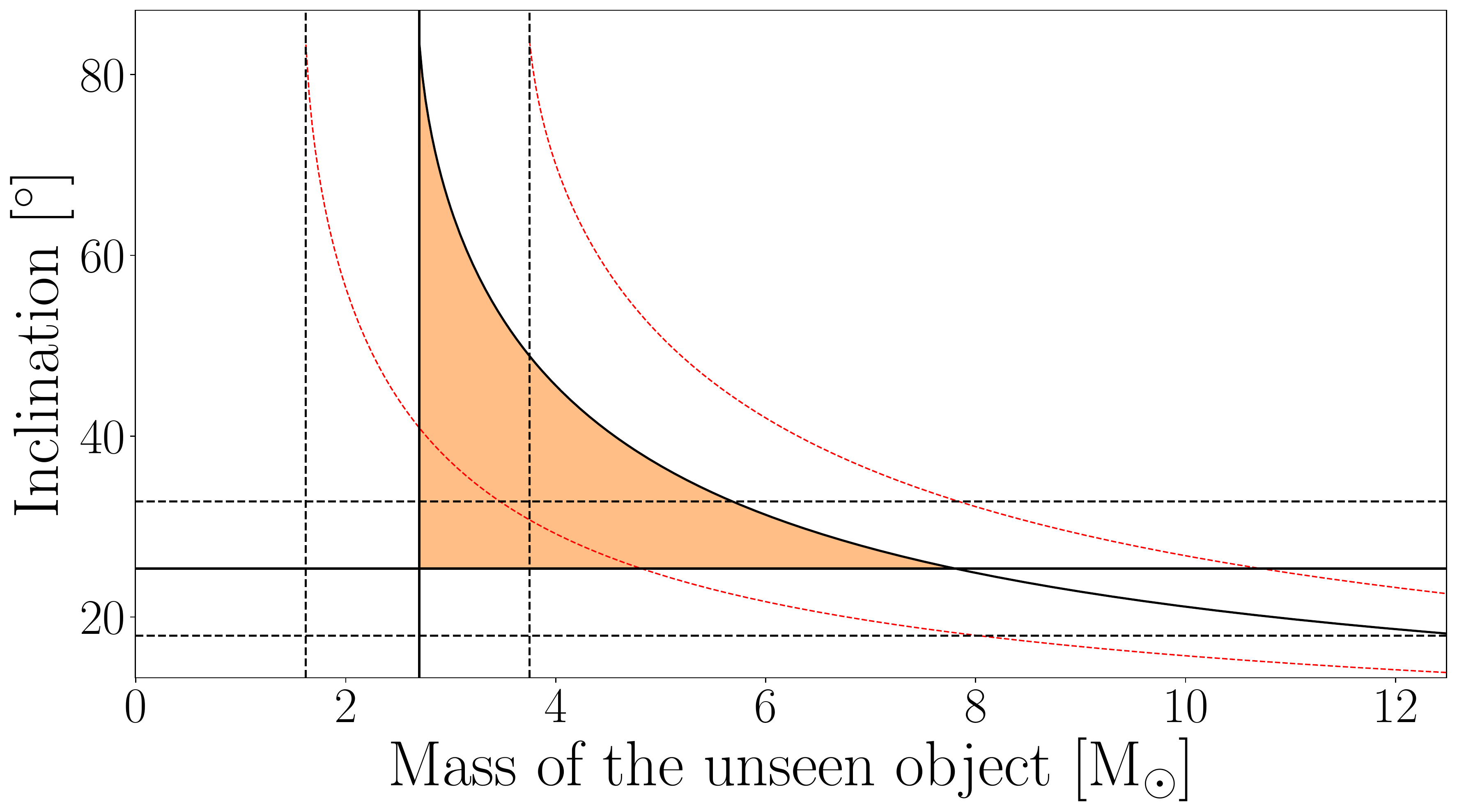}
            \caption{V* V946 Sco}
        \end{subfigure}
        \hfill
        \begin{subfigure}[b]{0.45\hsize}
            \centering
            \includegraphics[width=\hsize]{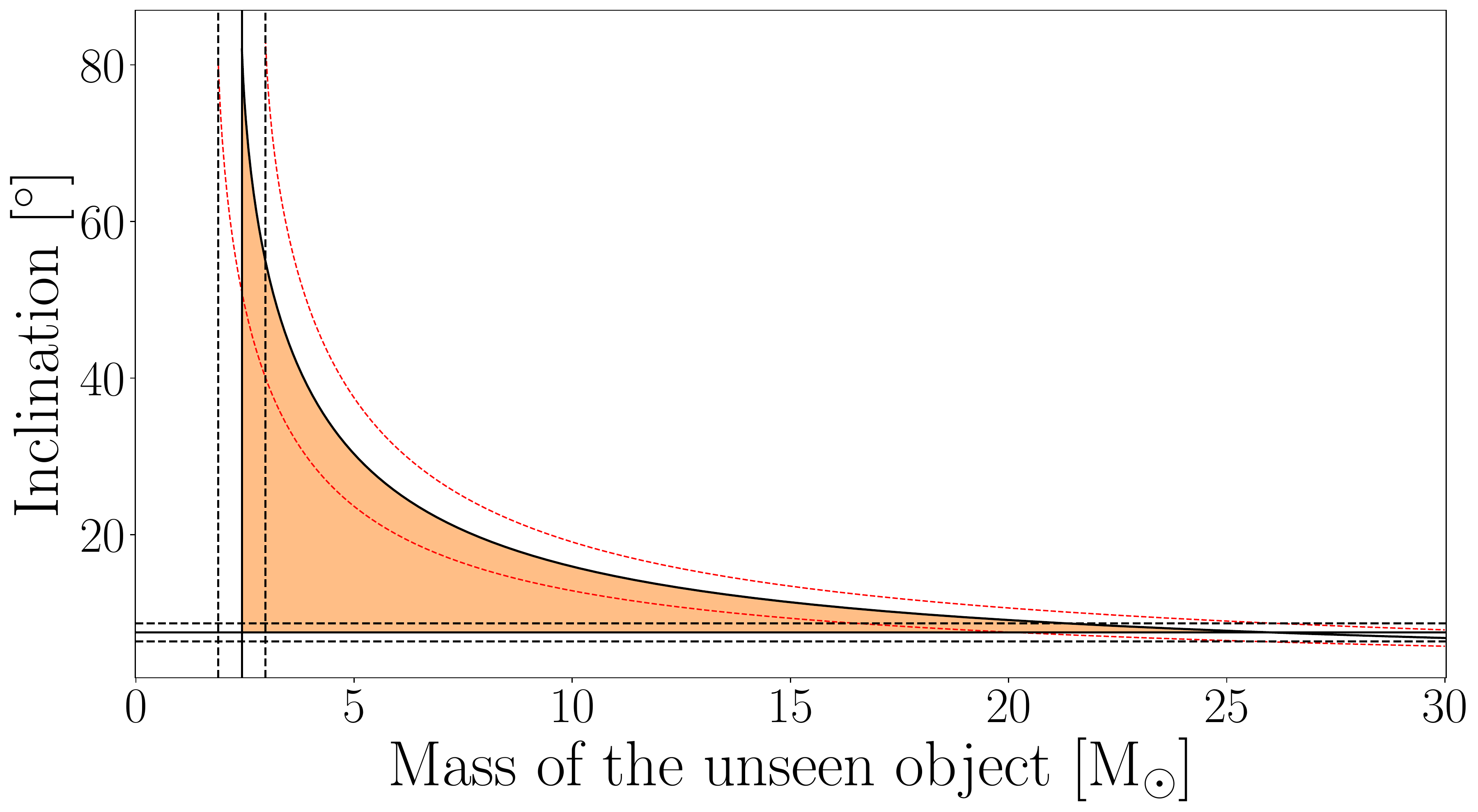}
            \caption{CD-41 11038}
        \end{subfigure}
            \hfill
        \begin{subfigure}[b]{0.45\hsize}
            \centering
            \includegraphics[width=\hsize]{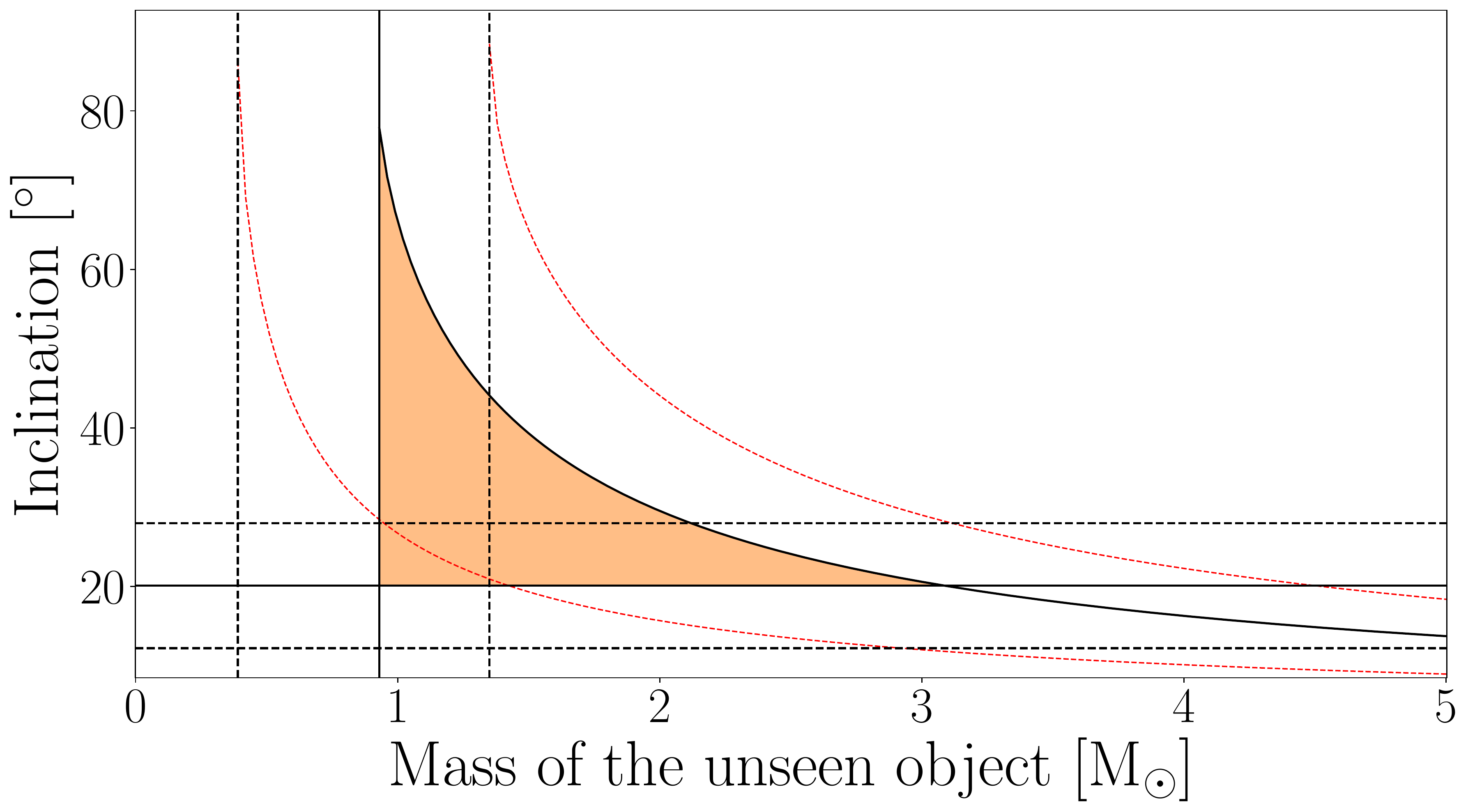}
            \caption{CXOU J165241.3-415536}
        \end{subfigure}
        \begin{subfigure}[b]{0.45\hsize}
            \centering
            \includegraphics[width=\hsize]{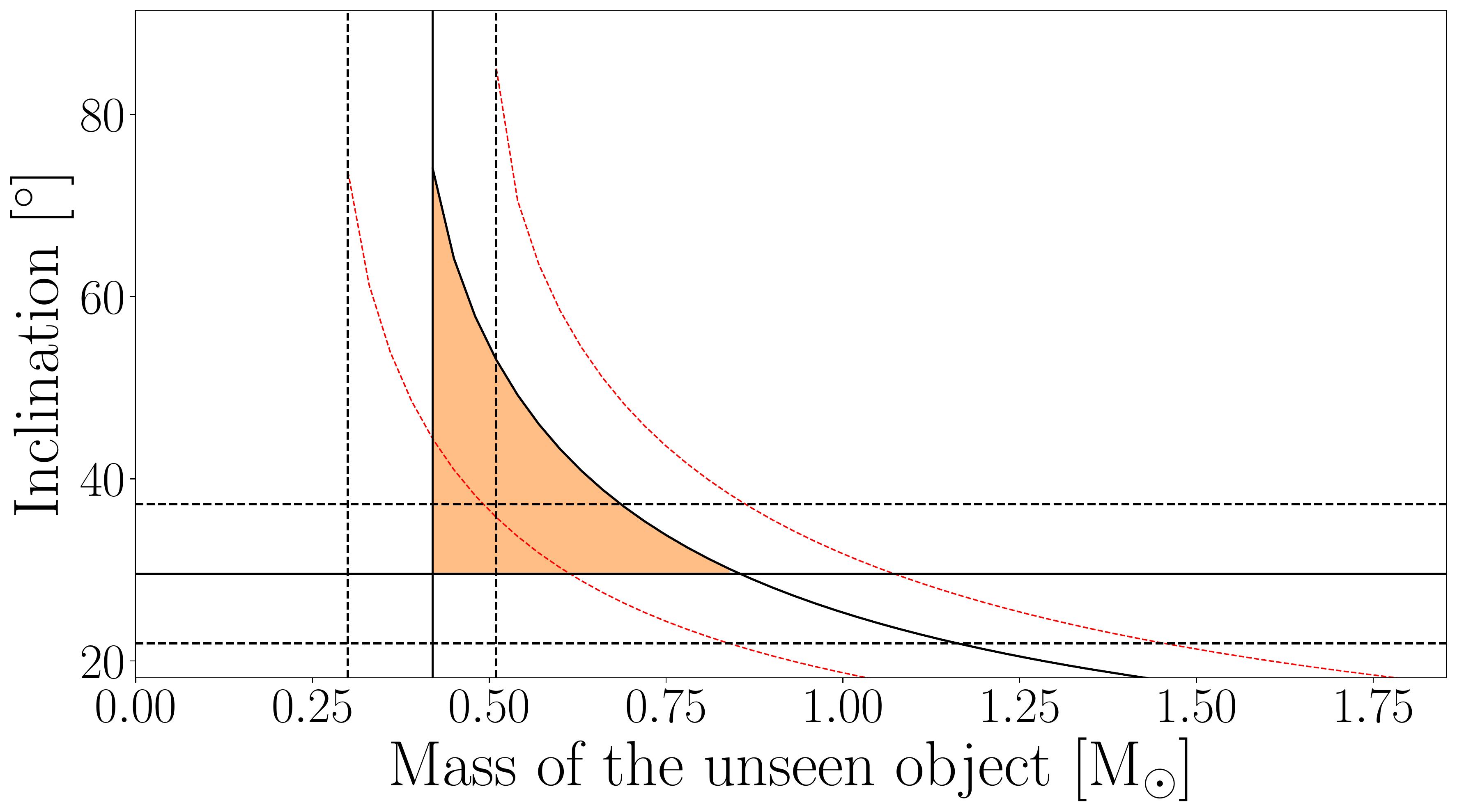}
            \caption{CPD-41 7717}
        \end{subfigure}
        \hfill
    \caption{The computed mass of the unseen companion of the SB1s as a function of the inclination of the system using Eqs. \ref{bmf2} and \ref{inclination}, and using the spectroscopic estimate of the the primary mass. The minimum inclination of each system is indicated by the horizontal line, with the associated error as dashed horizontal lines. The vertical line indicates the minimum mass of the unseen companions, with the associated error as dashed vertical lines. The orange shaded regions correspond to the possible values for the system inclinations and the masses of the unseen objects. The red dashed line corresponds to the uncertainty on the binary mass function, computed by propagating the 1$\sigma$ errors on the other parameters.}
    \label{fig:specmassinclination}
\end{figure*}

\begin{figure*}
    \centering
        \begin{subfigure}[b]{0.45\hsize}
            \centering
            \includegraphics[width=\hsize]{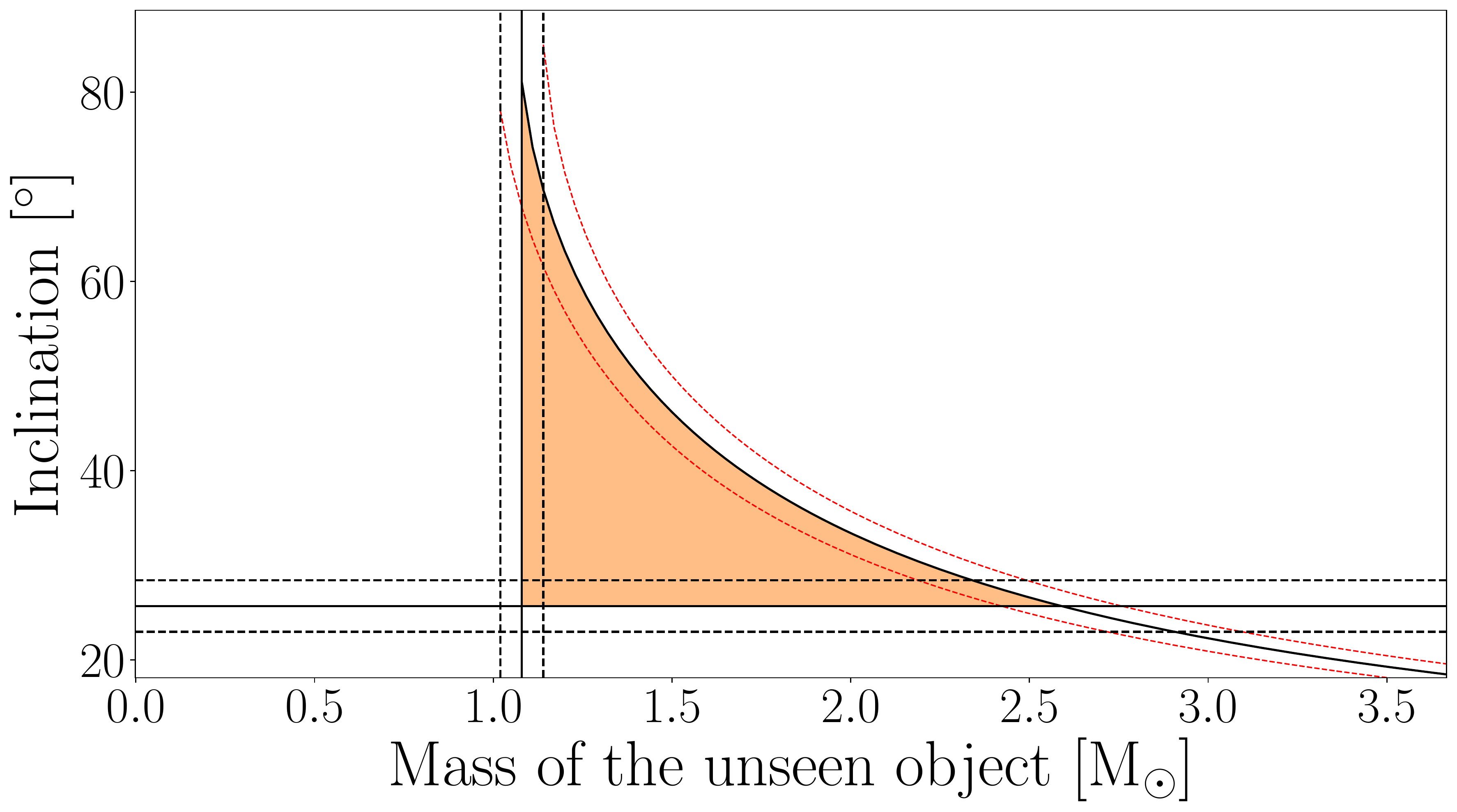}
            \caption{HD 152200}
        \end{subfigure}
        \hfill
        \begin{subfigure}[b]{0.45\hsize}
            \centering
            \includegraphics[width=\hsize]{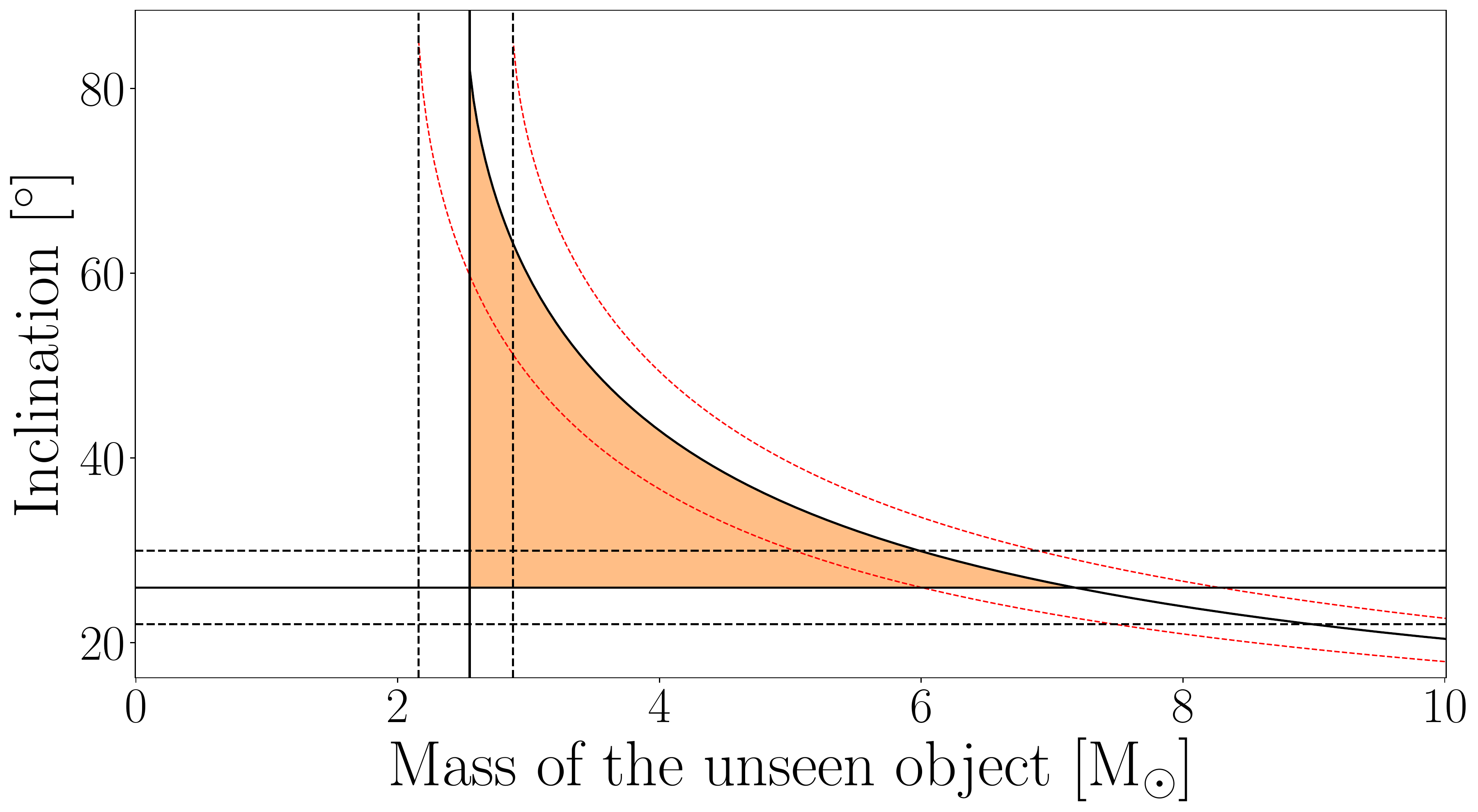}
            \caption{V* V946 Sco}
        \end{subfigure}
        \hfill
        \begin{subfigure}[b]{0.45\hsize}
            \centering
            \includegraphics[width=\hsize]{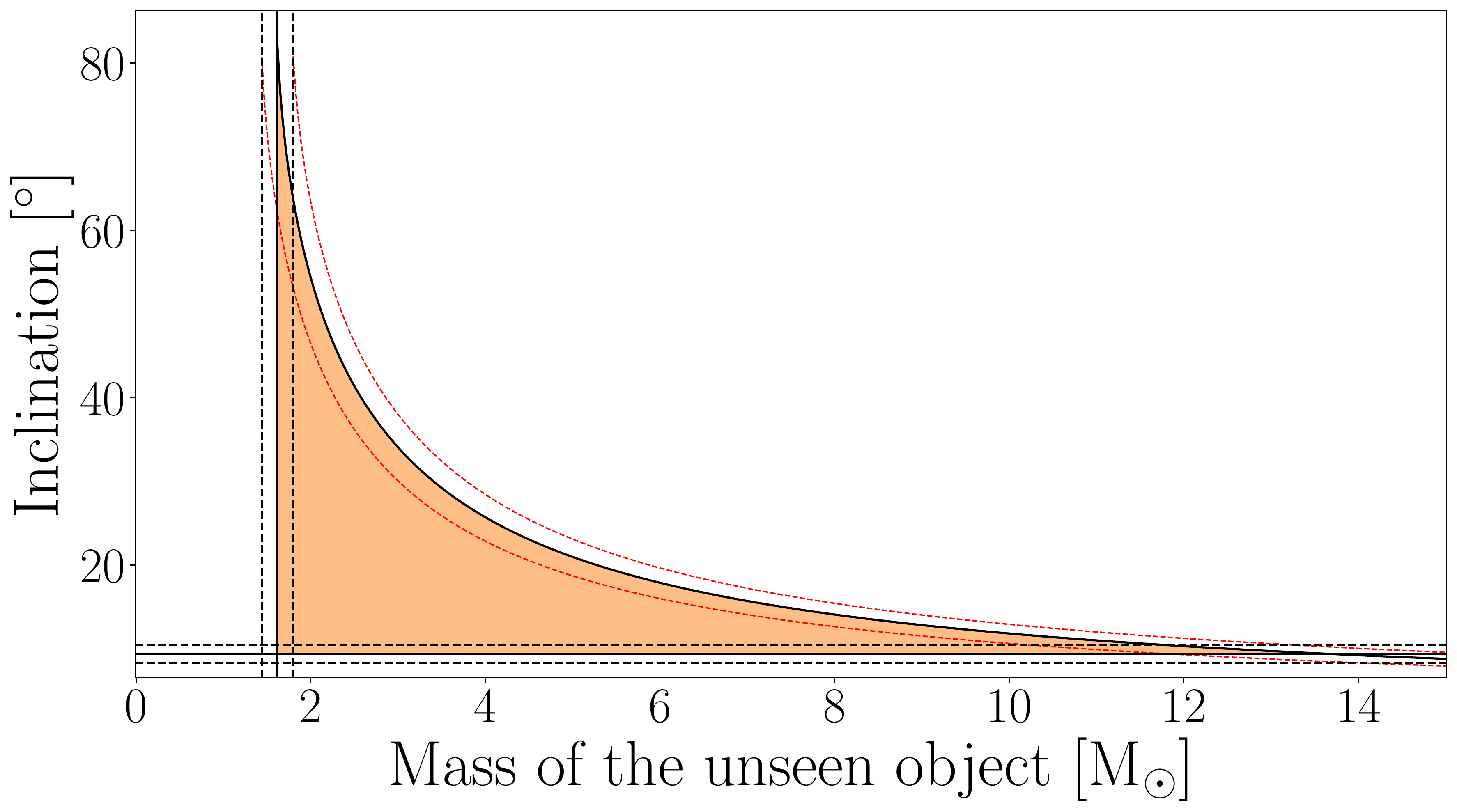}
            \caption{CD-41 11038}
        \end{subfigure}
            \hfill
        \begin{subfigure}[b]{0.45\hsize}
            \centering
            \includegraphics[width=\hsize]{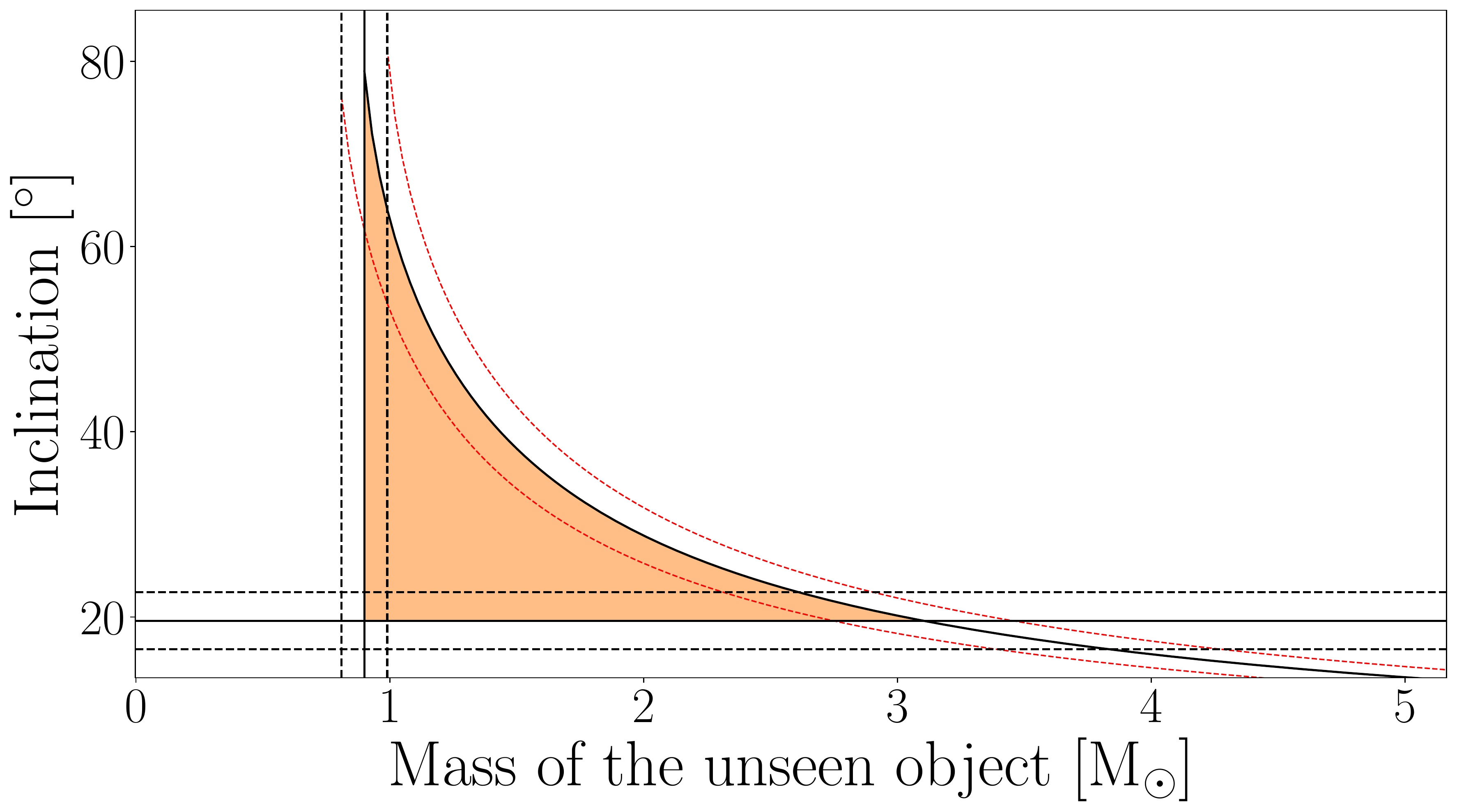}
            \caption{CXOU J165241.3-415536}
        \end{subfigure}
        \begin{subfigure}[b]{0.45\hsize}
            \centering
            \includegraphics[width=\hsize]{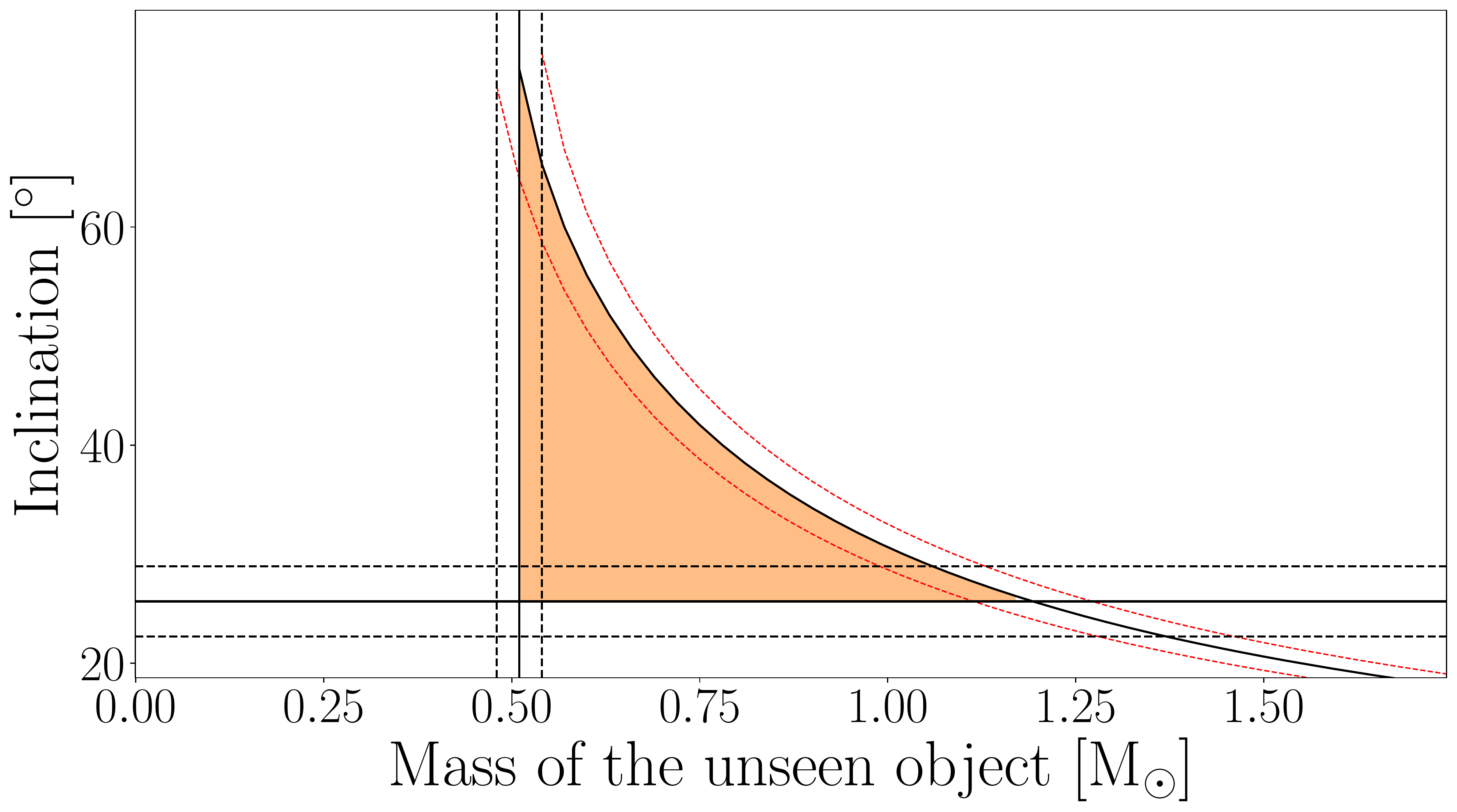}
            \caption{CPD-41 7717}
        \end{subfigure}
        \hfill
    \caption{Same as Figure~\ref{fig:specmassinclination}, but using the evolutionary estimate of the the primary mass.}
    \label{fig:evomassinclination}
\end{figure*}

 
We use Equations \ref{bmf2} and \ref{inclination}, which rely exclusively on the constrained orbital parameters (Paper~I) and stellar parameters (Section~\ref{ss:atmospheremodelling}), to compute the possible range of masses of the unseen companions of the candidate systems identified in Section~\ref{ss:disentangling}. Figure~\ref{fig:specmassinclination} shows the relation between the inclination of the systems and the mass of the unseen companion using the spectroscopic estimate of the primary masses, and Figure~\ref{fig:evomassinclination} shows the same, but calculated with the evolutionary estimate of the primary masses. \par

\subsection{TESS photometry}
\label{ss:TESS}

Another tool at our disposal in detecting lower mass stellar companions to the SB1s is photometry. For example, if one of these systems is found to be an eclipsing binary (EB), then logically the presence of stellar companion is inferred. However, there are also signatures of detached (and thus quiet, non-interacting) stellar BHs that can be found in space-based photometry as obtained by, for example, TESS. \citet{masuda_prospects_2019} have illustrated multiple potential periodic signals from the visible component in TESS light curves (LCs). Ellipsoidal variations occur as a result of tidal distortion of the primary star due to the gravity of the BH (and due to non-BH companions), causing the geometric shape and the brightness distribution of the visible star to change. Doppler beaming results in an amplitude variation of the light curve caused by the relativistic aberration of the visible star's light, time dilation, and the Doppler motion of the visible star's light. These two effects are dependent on and are in phase with the orbital period of the binary system. The third signature then, is self-lensing: this occurs when the BH eclipses the visible star, and the BH acts as a lens, gravitationally magnifying the star. It is, however, unlikely to find any indication in the light curve of the presence of a non-degenerate companion in systems with longer orbital periods. \par

\begin{figure*}
    \centering
        \begin{subfigure}[b]{0.45\hsize}
            \centering
            \includegraphics[width=\hsize]{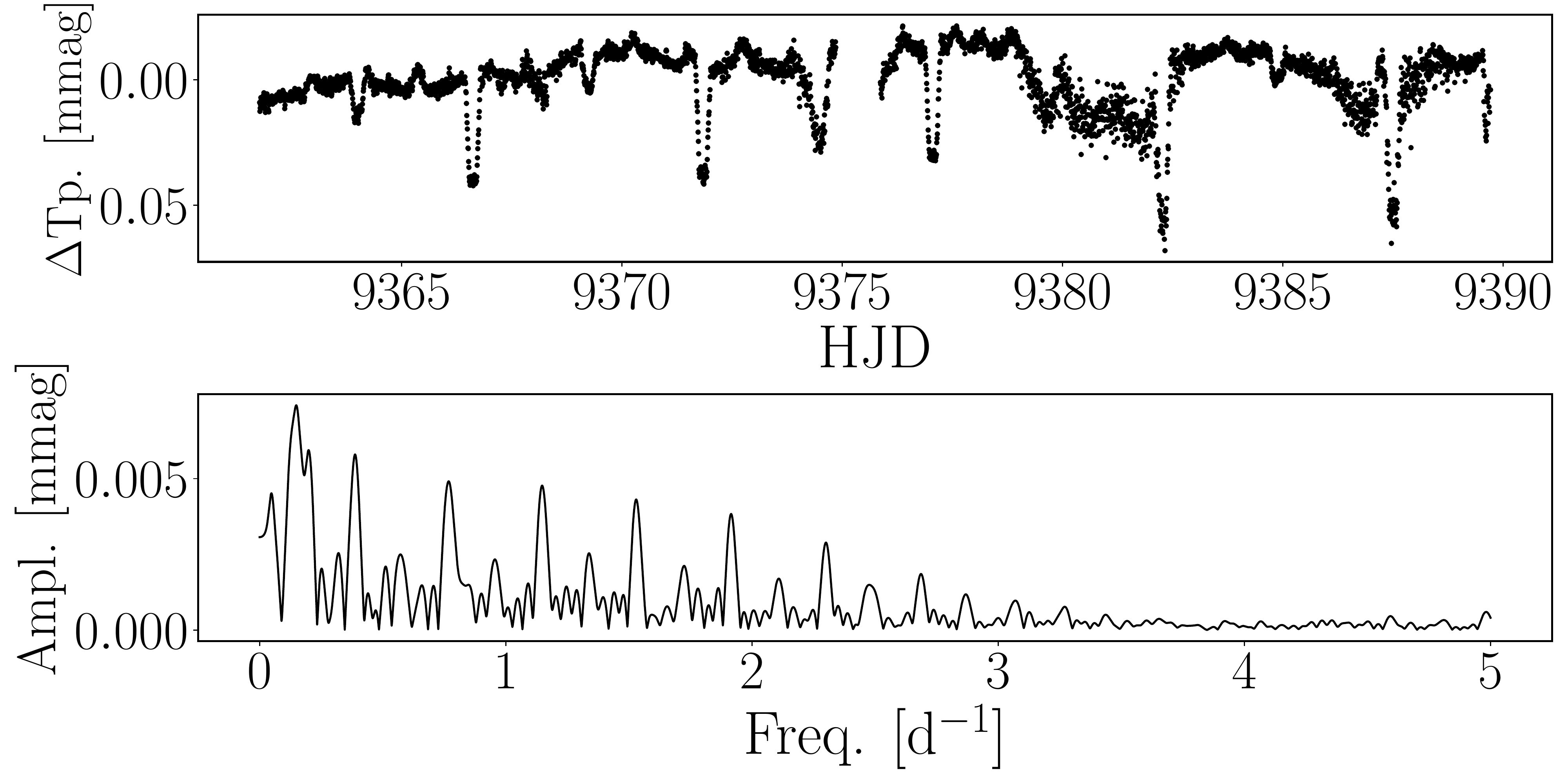}
            \caption{V* V1208 Sco}
        \end{subfigure}
        \hfill
        \begin{subfigure}[b]{0.45\hsize}
            \centering
            \includegraphics[width=\hsize]{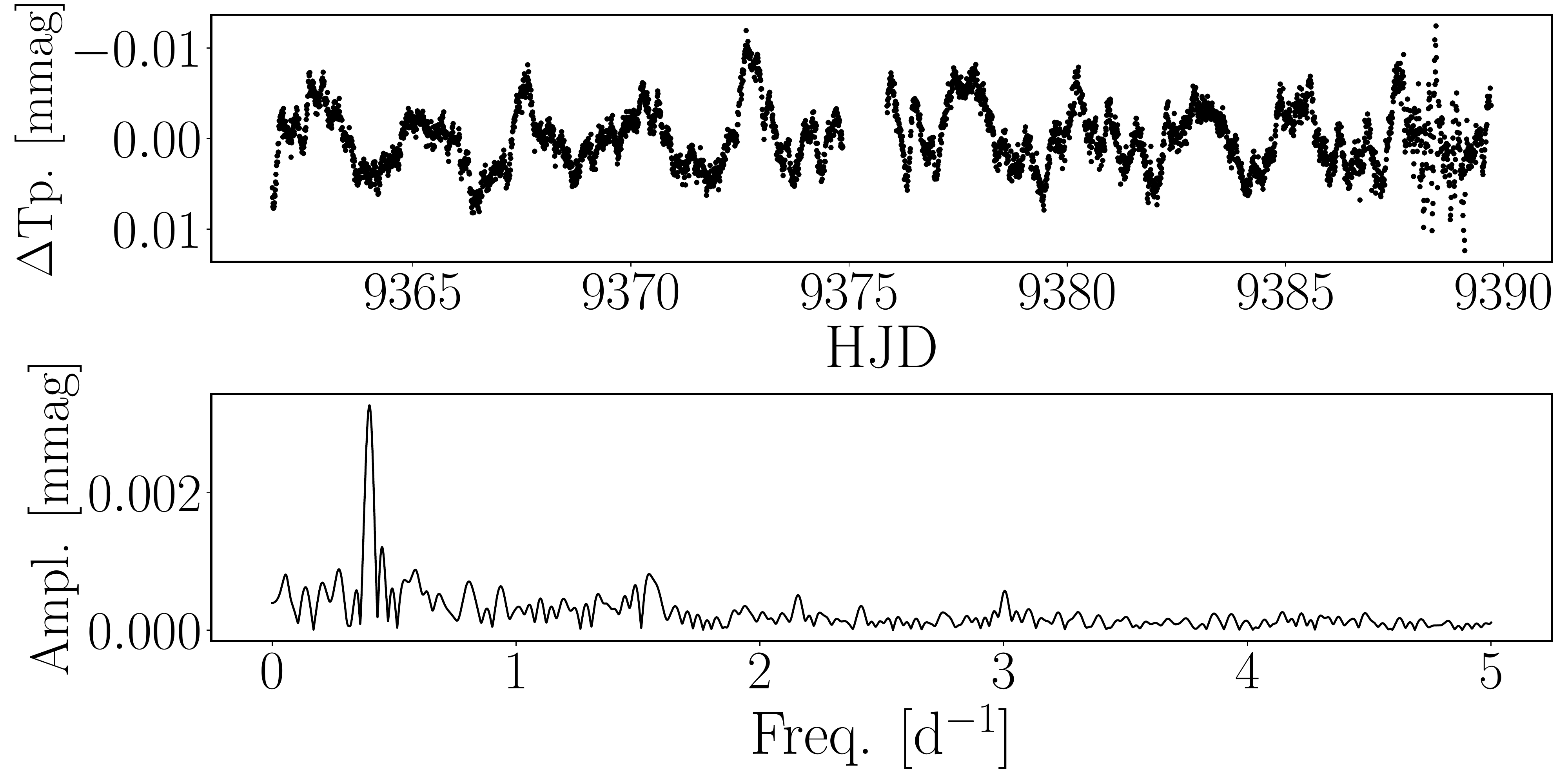}
            \caption{CPD-41 7717}
        \end{subfigure}
        \hfill
        \begin{subfigure}[b]{0.45\hsize}
            \centering
            \includegraphics[width=\hsize]{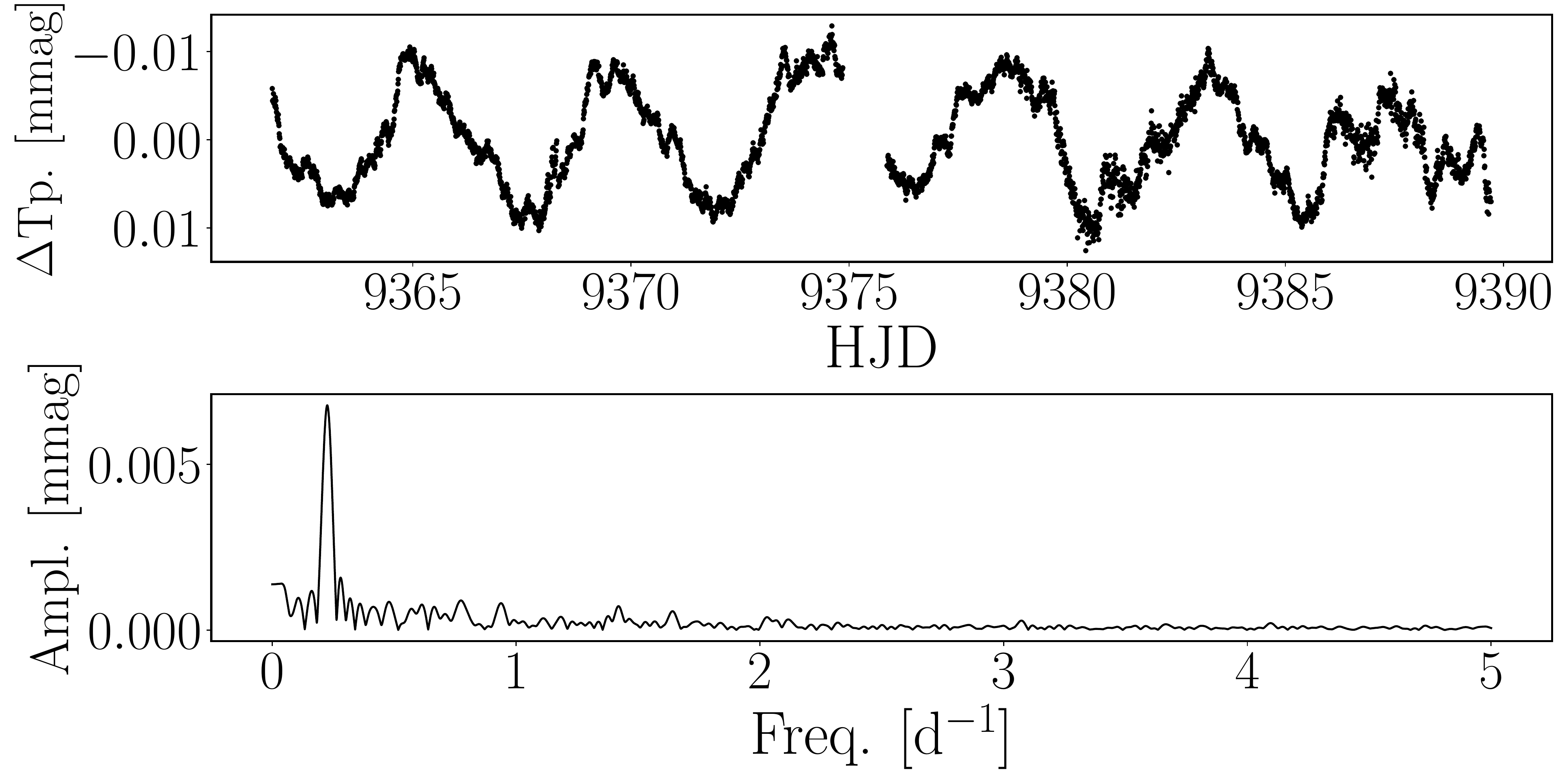}
            \caption{HD 152200}
        \end{subfigure}
            \hfill
        \begin{subfigure}[b]{0.45\hsize}
            \centering
            \includegraphics[width=\hsize]{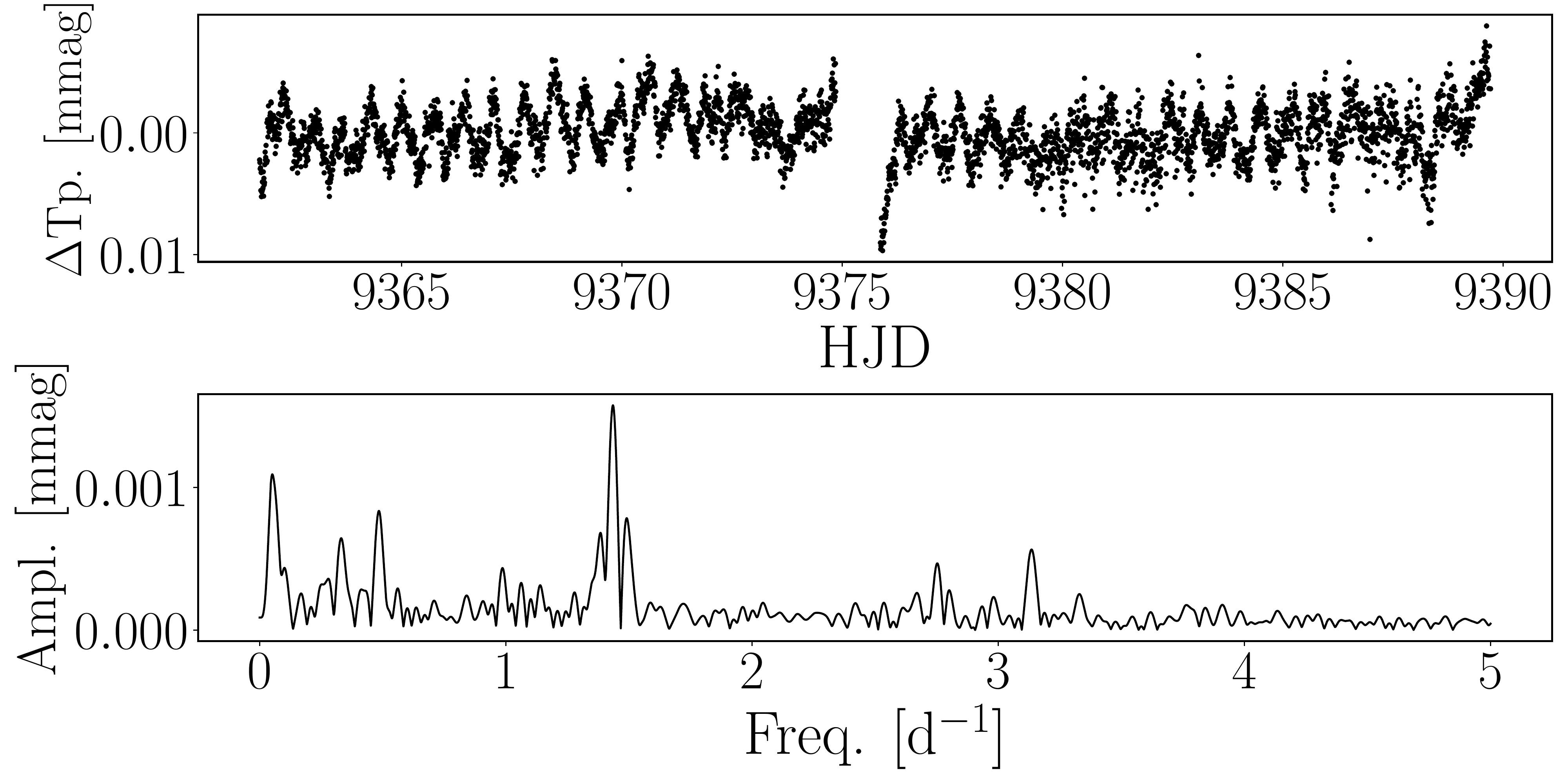}
            \caption{CXOU J165421.3-415536}
        \end{subfigure}
        \begin{subfigure}[b]{0.45\hsize}
            \centering
            \includegraphics[width=\hsize]{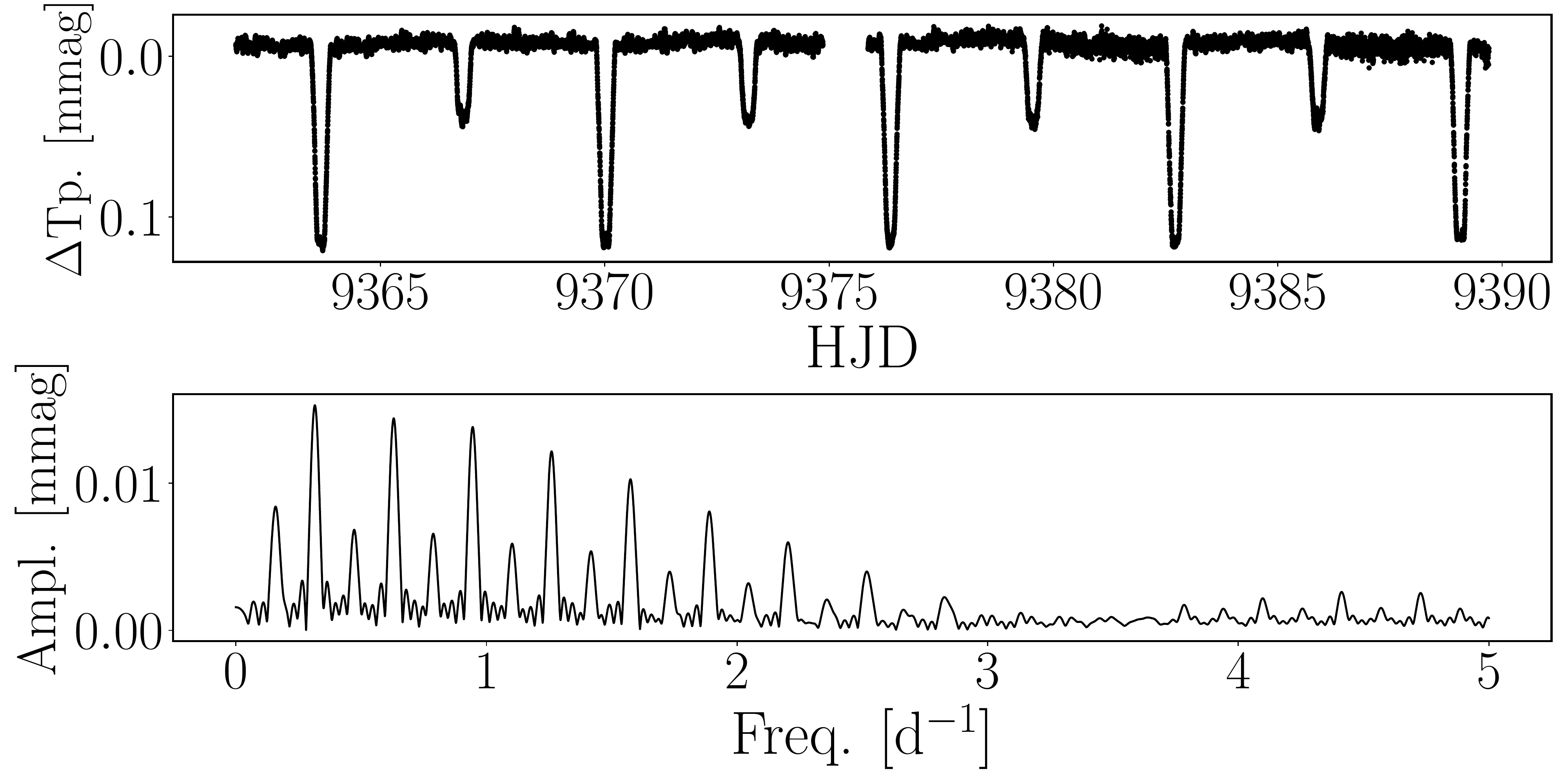}
            \caption{CPD-41 7746}
        \end{subfigure}
        \hfill
    \caption{TESS light curves and their periodograms of five of the targets in which we detected photometric variability.}
    \label{fig:tessLCs}
\end{figure*}

The retrieval and reduction of the TESS lightcurves is described in Section \ref{sss:photometry}. The light curves were observed in the TESS sector 39. We display in Fig.~\ref{fig:tessLCs} the light curves and their periodograms. These periodograms were computed using the Heck-Manfroid-Mersch technique \citep[][revised by \citealt{gosset_first_2001}]{heck_period_1985}. Because the field is very crowded, the extraction of the TESS light curves was challenging. We were not able to extract the TESS light curves for 6 objects in our SB1 sample: CD~$-41$~11030, CD~$-41$~11038, NGC~6231~78, NGC~6231~255, NGC~6231~273, and NGC~6231~225). All these objects suffer from contamination from other stars in their neighborhood. \par 

We found two systems showing eclipses in their light curves: V$^{*}$~V1208 Sco, and CPD-41 7746. These systems were already detected as SB2 after applying spectral disentangling (see Sect.~\ref{ss:disentangling}). Two other objects show photometric variability: HD 152200, and CPD-41 7717. We were not able to extract the spectra of the companions in these two objects after the spectral disentangling. The signals in these light curves have the same period as their orbital ones. This excludes ellipsoidal variations or eclipses as possible cause, where the signal would have had half the orbital periods. Given the projected rotational velocities, the estimated radii of the visible stars, and the detected periods, it is unlikely that these signals can come from rotation. Finally, the light curve of CXOU${\sim}$J165421.3$-$415536 has a significant frequency that corresponds to a period of 0.6967 days. If this signal is due to the rotation, that implies that the inclination of the star (and perhaps the system) is about 31$^{\circ}$. That inclination, assuming that the rotational axes are perpendicular to the orbital plane, would give a mass of about 1.8~$M_{\odot}$ for the unseen companion. However, we do not exclude that the light curve of CXOU~J165421.3$-$415536 might be affected by contamination from other close-by objects. \par

Previous photometric studies on some of the stars in our sample do not indicate any of the above signatures of BHs, but there are findings of note. \cite{meingast_pulsating_2013} performed a study with Johnson $UBV$ photometry to identify pulsating and otherwise variable stars in NGC 6231. Out of the initial 15 stars described in this work (both SB1s and newly double-lined SB2s), HD 326328 was found to be an slowly pulsating B-star (SPB) candidate, NGC 6231 225 was classified as an SPB, and V* V946 Sco was classified as a $\beta$ Cepheid star. Out of these three, only V* V946 Sco has been found to be a potential candidate for having a compact companion in this work. Otherwise, the only other star in the sample studied photometrically is HD 152200. \citep{pozo_nunez_survey_2019}, in a survey for high-mass eclipsing binaries in Sloan r and i photometry, found HD 152200 to be an eclipsing binary with a period of $P = 8.89365 \pm 0.00081$ days. This is almost exactly twice the orbital period found in Paper~I for this system, which was $P = 4.4440 \pm 0.0007$ days. It should be noted that the authors were unable to model the Roche geometry of the system using the obtained LC. If this system is indeed an EB, then this is not compatible with the proposal that the unseen companion is a compact object. The period is compatible with that of a slowly pulsating B-type star, however, the star's LC does not appear to indicate that the variability is solely due to pulsations. \par 

\section{Discussion}
\label{s:discussion}

\subsection{Newly identified SB2 systems}
\label{ss:sb2s}

\begin{figure}
\centering
\includegraphics[width=\hsize]{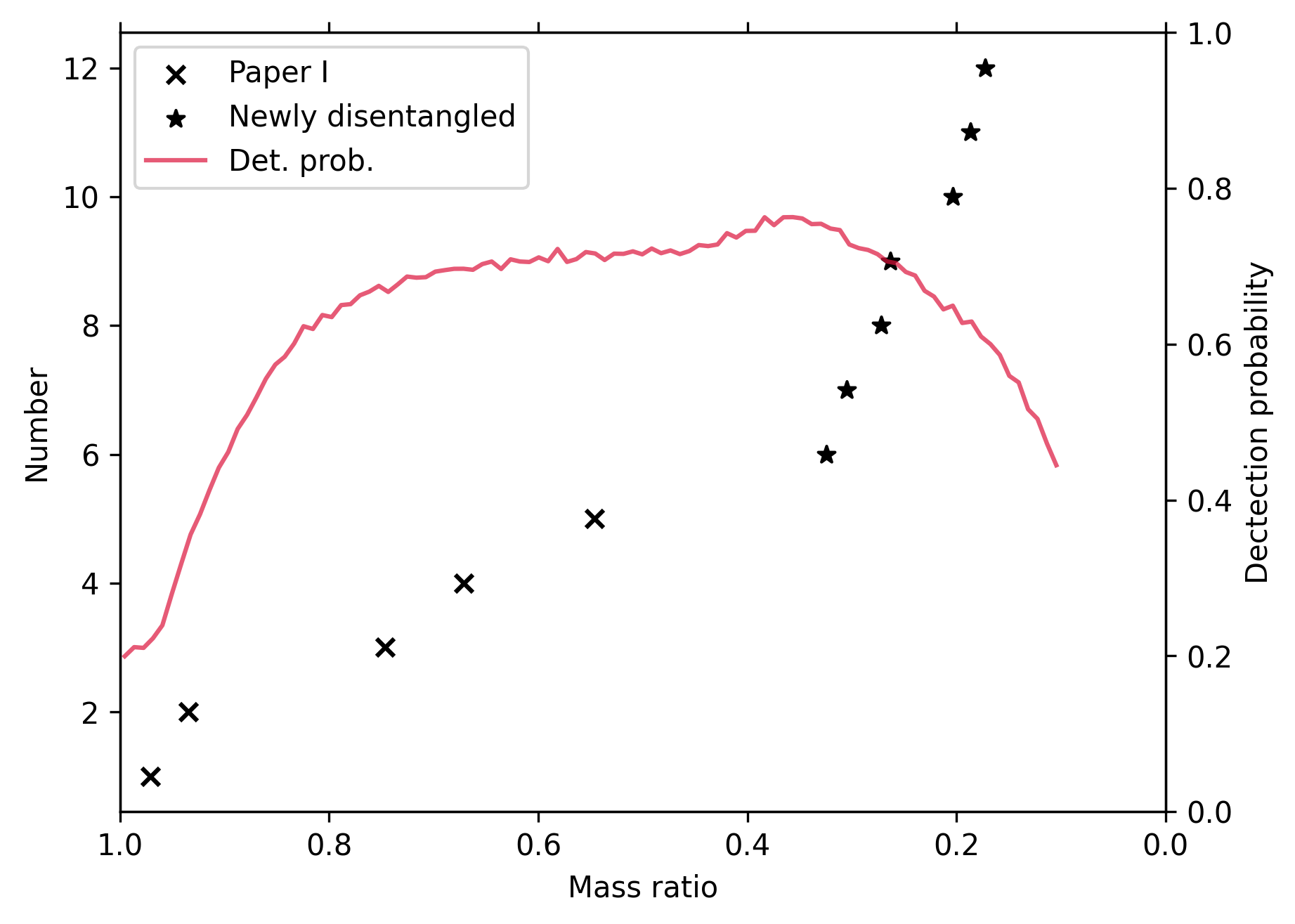}
    \caption{The observed cumulative mass-ratio distribution of the SB2 stars in NGC 6231, with the Paper~I SB2s as crosses, and the newly disentangled SB2s as stars. The detection probability, calculated from the bias correction of the observational campaign in Paper~I, is displayed as a red curve.  
         }
    \label{f:massratios}
\end{figure}

We found seven of the 15 previously identified SB1 B-type binaries of NGC 6231 to be SB2s through Fourier disentangling (Sec.~\ref{ss:disentangling}), where the signature of a secondary star has been extracted from the composite spectrum. These targets are NGC 6231 723, HD 326328, CD-41 11030, NGC 6231 78, V*V1208 Sco, CPD-41 7722, and CPD-41 7746 (see Fig.~\ref{FigSB2Spec} for the disentangled spectra). We have been able to probe down to a mass ratio of near 0.1, which is significantly lower than what was found in Paper~I. Fig.~\ref{f:massratios} shows the mass ratio distribution of the NGC 6231 B-type SB2s, along with the detection probability curve computed for the observational campaign to account for the observational biases. The original campaign had a significant drop in sensitivity to objects with a mass-ratio below 0.4, and as was proposed in Paper~I, most of the missing low mass ratio systems have been found in the originally proposed "SB1s". \par
\begin{figure}
\centering
\includegraphics[width=\hsize]{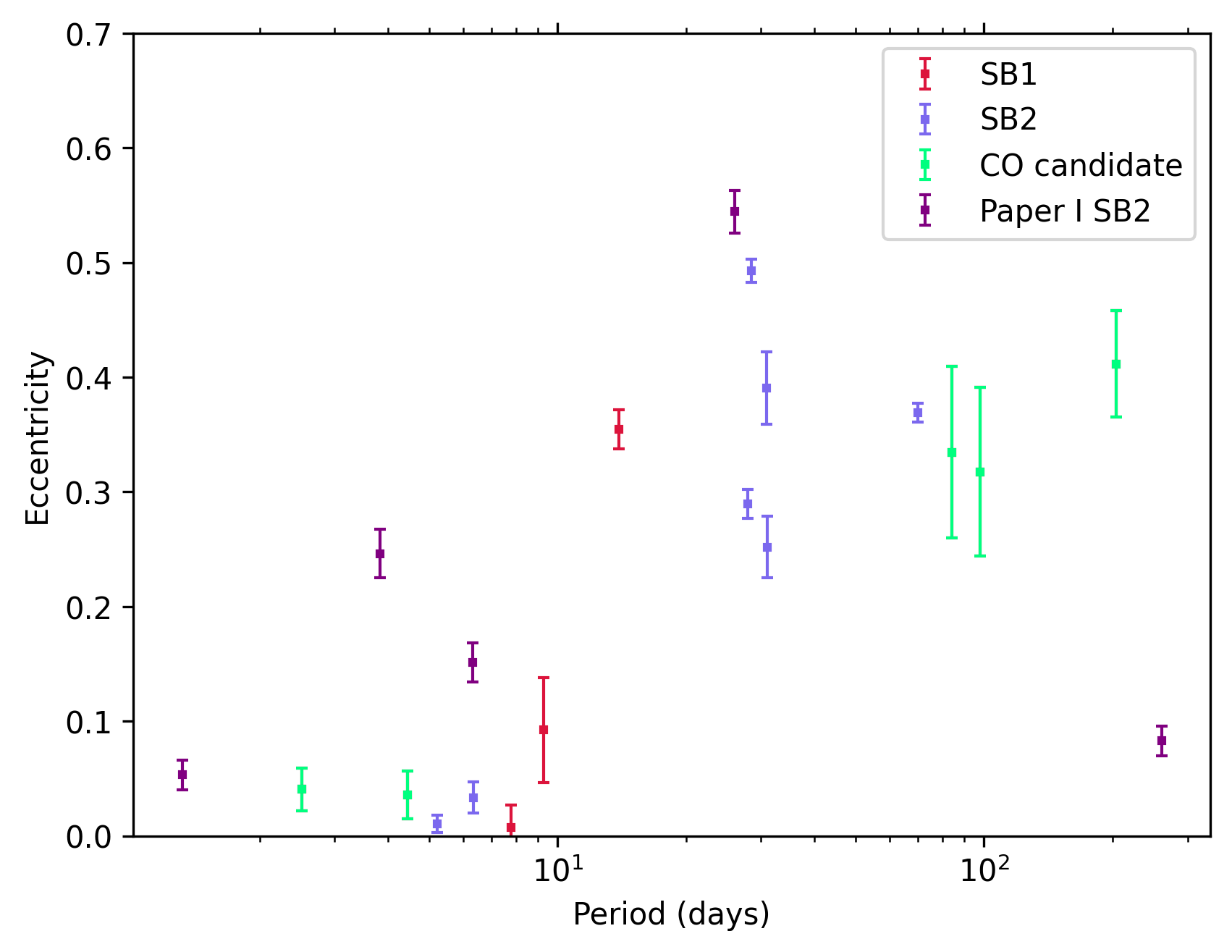}
    \caption{Orbital period against eccentricity of the B-type binaries in NGC 6231. The SB1s are shown in red, with the candidates for harbouring compact objects in green, the Fourier disentangled SB2s in blue, and the SB2s identified in Paper I in purple.}
    \label{f:pve}
\end{figure}

 
Figure~\ref{f:pve} shows the orbital periods against the eccentricities of the newly disentangled systems, alongside the SB1s and the SB2s identified visually in the Paper~I. Most of these SB2 systems seem to have periods around 30 days, and this was previously noted as a feature in the full period distribution (SB1s and SB2s) in Paper~I, and could not be explained by alliasing of the frequency space of the observational campaign by, for example, its cadence. We also found the new SB2s to have eccentricities from effectively circular up to around 0.5.

\subsection{Candidates for harbouring compact objects}
\label{ss:COs}

\begin{figure*}
\centering
\includegraphics[width=\hsize]{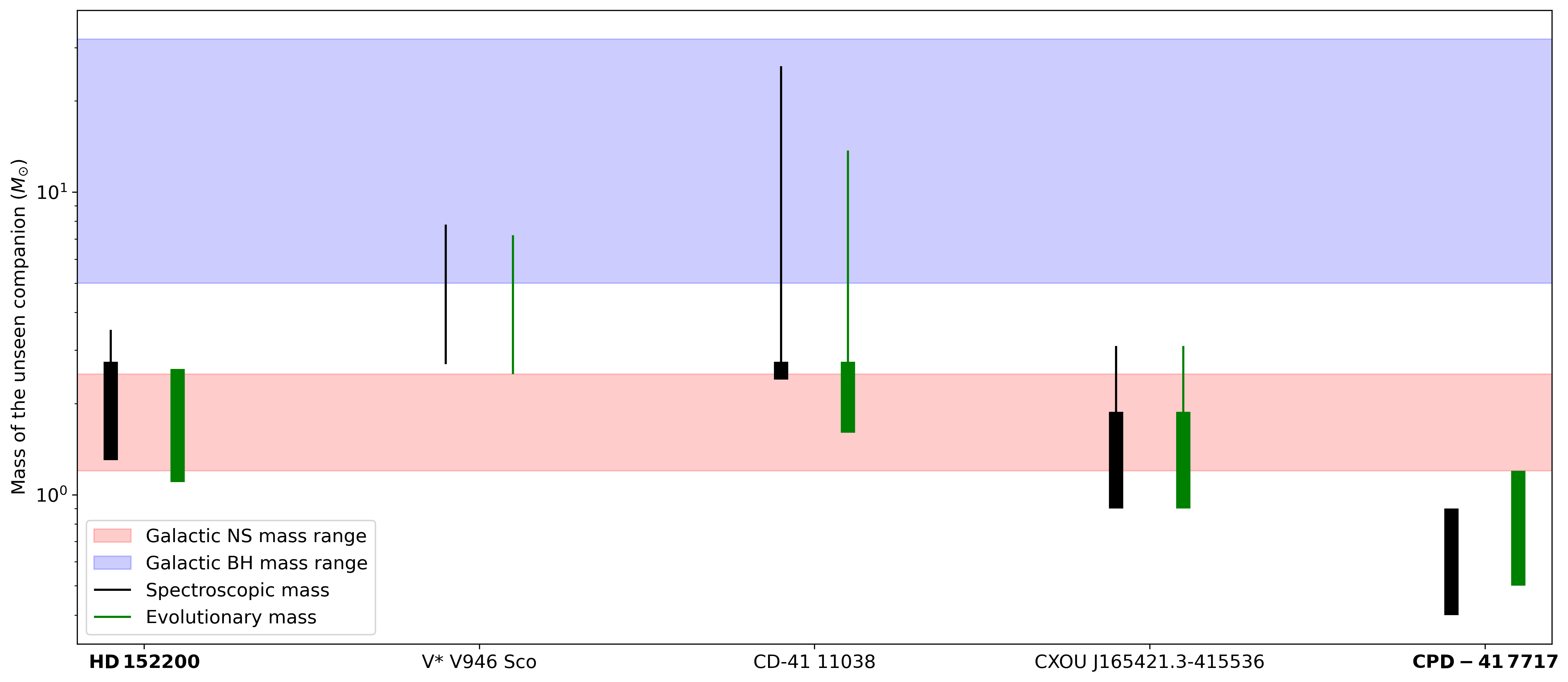}
    \caption{Estimated spectroscopic (black) and evolutionary (green) masses of each SB1's unseen companion. The blue zone corresponds to the predicted mass range for Galactic stellar mass BHs and the red zone represents the same but for Galactic NSes \citep{belczynski_maximum_2010,fryer_hypercritical_2014}. The mass ranges for each target where main sequence stars cannot be discounted according to the spectral disentangling sensitivity (Section \ref{ss:disentangling}) are denoted by thick lines. Target names in boldface are the targets for which no companion could be extracted through spectral disentangling, and the other targets are those for which the nature of the extracted companion was ambiguous (Sect.~\ref{ss:disentangling}).  
         }
    \label{f:cocandidates}
\end{figure*}

The remaining eight targets, then, are those that we cannot extract a companion's signature through Fourier disentangling of their spectra. As stated in Sect. \ref{sss:spectro_fit}, three of these have primary stars of spectral type B9 and very small binary mass functions, and so we exclude them from the candidates for harbouring compact objects. This leaves a remaining five candidate systems. Figure~\ref{f:cocandidates} shows both the estimated spectroscopic and evolutionary estimates of each candidate system's unseen companion, as calculated with the the binary mass function (Eqn. \ref{bmf2}) with the estimated spectroscopic (Sec.~\ref{ss:atmospheremodelling}) and evolutionary masses (Sec.~\ref{ss:evofitting}). Also shown in Figure~\ref{f:cocandidates} is the predicted mass range for Galactic stellar BHs and NSes \citep{belczynski_maximum_2010,fryer_hypercritical_2014}. For the NSes, this is between 1.2--2.5~$M_{\odot}$, and for BHes this is above 5~$M_{\odot}$.  \par 

\subsubsection{HD 152200}
\label{sss:hd152200}

HD 152200 was spectrally classified as a B0V star with a 4.44~d orbital period with circular eccentricity in Paper~I. In this work, the star's spectroscopic mass was constrained to be 23.4 $\substack{+15.6 \\ -13.6}~M_{\odot}$, and the evolutionary mass constrained was 17.8 $\substack{+1.2 \\ -0.8}~M_{\odot}$. Using the binary mass function, and the constrained primary mass estimates and the orbital parameters of the system, the range of possible masses for the companion of HD 152200 is between 1.1 and 3.5 $M_{\odot}$ (analogous to a late F to a B8 MS star). \par 

During disentangling, this star was classified as a target from which we could not unambiguously extract a companion from the spectra. According to the simulations described in Section \ref{ss:disentangling}, which were designed to reject the presence of non-degenerate companions down to certain mass and flux ratios, it was found that, considering a B0V primary, we are able to extract the spectrum of a secondary object down to a spectral type of B9V or A0V. We can likely reject the possibility of a hidden non-degenerate MS companion above 2.2 $M_{\odot}$, but this does not discount the entire range of possible masses for HD 152200's companion, for which the minimum is 1.1 $M_{\odot}$. When considering a stripped star companion, we can expect to be able to extract a stripped star companion with a temperature $T_{\rm eff} = 62$~kK and a radius of $R \sim 0.55~R_{\odot}$, i.e., an object with an initial mass of about $M_{\rm ini} \sim 8.5~M_{\odot}$ and a current stripped mass of $M_{\rm strip} \sim 2~M_{\odot}$. As we have limited information due to the limited wavelength space of our composite spectrum, we are missing access to signatures of a stripped star companion, so this does increase the difficulty in detecting these stars. One of these, for example, is strong \ion{He}{ii}~$\lambda$4686 emission. The wavelength region we have access to, 3964--4567\AA, mostly consists of weak absorption lines. However, the most massive stripped stars in the \cite{gotberg_spectral_2018} models, i.e. the WN3 stars (those with $M_{ini} > 15~M_{\odot}$), have strong emission features in the wavelength domain we are observing, and so it is likely we can exclude these. Note, however, that the strengths of the line depend on the assumed wind mass loss rates which are uncertain for strepped stars. As for the possibility of being a triple system, it would be unlikely due to HD 152200's short period. Based on the prediction of the unseen masses, if the companion was a compact object, it would likely be a neutron star. \par

Regarding the light curves of HD 152200, the target remains an ambiguous case with reports of the star being an eclipsing binary with a period of twice that found in Paper~I \citep[$P = 8.89365 \pm 0.00081$~d, ][]{pozo_nunez_survey_2019}. While there is definitely photometric variability, there do not appear to be eclipses in the TESS photometry (see Fig.~\ref{fig:tessLCs}). As mentioned in Sec.~\ref{ss:TESS}, as the variability appears to have the same period as the orbital period constrained in Paper~I, this excludes ellipsoidal variations (due to a close compact object companion, for example) or eclipses. This remains an interesting object that would benefit from further photometric study. \par

The stellar content of NGC 6231 has been subject to multiple X-ray surveys in the past. An \textit{XMM-Newton} campaign towards NGC 6231 was performed by \cite{sana_xmm-newton_2006}, and \textit{Chandra} observations were obtained by \cite{damiani_chandra_2016} for the stars of NGC 6231. HD 152200 appears across the \textit{XMM-Newton} and \textit{Chandra} surveys, and the surveys find HD 152200 to have an X-ray luminosity of log$L_X = 31.29$ and log$L_X = 31.32$, respectively. Given the early-type nature of these stars, this value is not unexpected and does not suggest active accretion from a compact object. Along with this, HD 152200 (and the rest of the CO candidates) are not detected in the Second ROSAT all-sky survey \citep{boller_second_2016}. This suggests the potential compact objects in these binary systems are X-ray quiet (dormant), but does not disqualify these companions from being lower-mass MS stars. \par

\subsubsection{V* V946 Sco}

V* V946 Sco was found to be a B2V star with a 97.9~d orbit in Paper~I. The star's spectroscopic and evolutionary masses were constrained to be 10.4 $\substack{+0.6 \\ -4.9}~M{\odot}$ and 9.4 $\substack{+0.6 \\ -0.4}~M_{\odot}$, respectively. The companion to V* V946 Sco, then, is estimated to be between 2.5 to 7.8 $M_{\odot}$ when taking both the spectroscopic and evolutionary primary mass estimates into account, which would be classified between B9 and B2. \par 

The disentangling status of this star was that the nature of the extracted companion was ambiguous. The detection limits we found for B2V stars were MS companions of spectral type down to A5V or A7V, corresponding to masses of 1.4 and 2.0~$M_{\odot}$. This means that if V* V946 Sco's companion was a MS star, we should be able to extract it from the spectra. If the companion was a stripped star, for a B2V primary star, we can extract the spectrum of a stripped star down to an $T_{\rm eff} = 43$~kK and a radius of $R \sim 0.34~R_{\odot}$, i.e., an object with an initial mass of about $M_{\rm ini} \sim 4.5~M_{\odot}$ and a current stripped mass of $M_{\rm strip} \sim 1~M_{\odot}$. Another case that was considered was a triple system, with two tight, low-mass stars forming the inner system and the B-type star in a wider orbit. The contribution of two F stars in the inner system, which would contain a total mass of $\sim 2-3~M_{\odot}$, would be very difficult to extract, and so this remains a possibility as this is within the range of possible masses for the companion of V* V946. However, two late-A stars in the inner system, with a combined mass of $\sim3.5~M_{\odot}$, should be detectable, and so we can exclude systems of these masses. If the two stars are any earlier than A7V, the system would likely become unstable. \par. 

As V* V946 Sco has a relatively long orbital period compared to the duration of a sector of TESS, it is not likely to have an indication in its light curve of the presence of a non-degenerate companion with TESS. It's orbital period is, however, close to the peak of predicted period distribution of OB+BH binaries, and also close to the peak of the observed orbital period distribution of Galactic Be/X-ray binaries \citep{langer_properties_2020}. This also applies to the targets CD-41 11038 and CXOU J165421.3-415536. \par 

Finally, the X-ray luminosities listed in the \textit{XMM-Newton} and \textit{Chandra} surveys are $\log L_X = 30.06$ and $\log L_X = 30.87$, respectively. Again, these are not unusual for non-accreting early-type stars. However, in these longer period cases, the low X-ray luminosity does not discount wind accretion by a potential compact object companion, as in a 100~d orbit, wind accretion does not result in detectable X-ray emission \citep{mahy_identifying_2022}.

\subsubsection{CD-41 11038}

CD-41 11038 was classified as a B0V star with a 83.9~d period. The spectroscopic and evolutionary masses constrained for the object are 32.1 $\substack{+3.8 \\ -7.4}~M{\odot}$ and 17.2 $\substack{+0.9 \\ -0.9}~M{\odot}$ respectively. It should be noted that the spectroscopic estimate of the mass far exceeds the mass expected of a B0V star, and corresponds more closely to that of a 07V/08V star. In Paper I, these stars were classified with the criteria in \citep{evans_vlt-flames_2015}, where the earliest classification for B-type dwarfs was B0V. Following these mass estimations, CD-41 11038 is estimated to have a companion of a mass between 1.6 and 26 $M_{\odot}$, and would be classified as a late-A/early-F star up to around spectral type O8, and also enters the predicted mass range for Galactic stellar BHs. \par

CD-41 11038 is another target for which the nature of the extracted companion was ambiguous. Again, like HD 152200 (also a B0V) star, we can expect to be able to extract MS companions down to spectral type A0V, meaning that we should be able to discount any MS companions above 2.2~$M_{\odot}$. However, the minimum possible mass of the unseen companion does undercut this slightly, so it is possible that an early-F/late-A star is present, yet hidden. If we consider a stripped companion, like HD 152200, we should be able to extract one with an initial mass of $M_{\rm ini} \sim 8.5~M_{\odot}$ and a current stripped mass of $M_{\rm strip} \sim 2~M_{\odot}$. As for the case of CD-41 11038 being in a triple system, we again cannot discount an inner binary system consisting of F-type stars, but we can exclude late A- and earlier-type inner systems as they would either be detected or dynamically unstable.. \par 

CD-41 11038's TESS lightcurve is unfortunately contaminated with light from neighbouring stars, so no information is able to be gleaned from it. However, it does also have a relatively long period compared to the duration of a TESS sector, and so is unlikely to show an indication in its light curve of the presence of a non-degenerate companion, as it would be unlikely that an eclipse would be observed. \par 

The X-ray luminosities for the star found by the \textit{XMM-Newton} and \textit{Chandra} surveys are log$L_X = 31.78$ and log$L_X = 31.57$ respectively. This is again typical for non-accreting early-type stars, but the same wind accretion possibility as for V* V946 Sco applies here. \par

\subsubsection{CXOU J165421.3-415536}

CXOU J165421.3-415536 was classified as a B2V star in Paper~I, with a 204~d period constrained for the SB1 system. The spectroscopic and evolutionary masses constrained for this target are 7.3 $\substack{+0.4 \\ -5.2}~M_{\odot}$ and 7.0 $\substack{+0.6 \\ -0.4}~M_{\odot}$ respectively. Consequently, the range of masses of the unseen companion is predicted to be between 0.9 and 3.1 $M_{\odot}$, which corresponds to a late-G to a B8 main sequence star. \par

CXOU J165421.3-415536 is the final target in our study for which the nature of the extracted companion was ambiguous. As CXOU J165421.3-415536 is a B2V star, according to our detection limit simulations, we should be able to extract the spectra of MS companions of spectral types down to A5V or A7V, corresponding to masses of 1.4 and 2.0~$M_{\odot}$, which does mean that there is potentially a solar-like or even sub-solar mass MS companion to CXOU J165421.3-415536 that we do not detect. As for a stripped star companion, we can expect to extract the spectrum of a stripped star down to an $T_{\rm eff} = 43$~kK and a radius of $R \sim 0.34~R_{\odot}$, i.e., an object with an initial mass of about $M_{\rm ini} \sim 4.5~M_{\odot}$ and a current stripped mass of $M_{\rm strip} \sim 1~M_{\odot}$. As for the triple system case, we again cannot discount an inner system consisting of two F stars, though we can disqualify late A- and earlier-type systems due to both being able to detect them and also to the instability of the earlier-type systems. \par 

The \textit{Chandra} survey provides an X-ray luminosity of log$L_X = 31.78$ and log$L_X = 30.16$, and which is again typical for quiescent early-type stars, but again we cannot discount the possibility of wind accretion from a compact object due to the system's relatively long orbital period.

\subsubsection{CPD-41 7717}

CPD-41 7717 was classified as a B2V star in Paper~I with an orbital period of 2.5~d. CPD-41 7717 is the last object amongst our candidates for which we could not unambiguously extract a companion. The spectroscopic and evolutionary masses constrained for the star are 6.1$\substack{+1.9 \\ -2.4}~M_{\odot}$ and 8.2$\substack{+0.6 \\ -0.3}~M_{\odot}$. The resulting range of possible unseen companion masses, 0.4 to 1.2$~M_{\odot}$, is below the predicted mass range for Galactic NSes, and it is likely that a solar-like companion is hidden in the composite spectrum. This mass range is also beneath the estimated detection limit for MS star companions, which for B2V stars is down to A5V or A7V stars, corresponding to masses of 1.4 to 2.0~$M_{\odot}$, so it is likely that the companion is a low mass MS star, below the detection limits. Regarding a potential stripped star companion, we can expect to extract the spectrum of a stripped star down to an $T_{\rm eff} = 43$~kK and a radius of $R \sim 0.34~R_{\odot}$, i.e., an object with an initial mass of about $M_{\rm ini} \sim 4.5~M_{\odot}$ and a current stripped mass of $M_{\rm strip} \sim 1~M_{\odot}$. \par

The short orbital period of CPD-41 7717 makes it viable to use TESS to reject a non-degenerate companion through the possibility of detecting an eclipse. CPD-41 7717 does show photometric variability in its LC, however, we do not observe any eclipses. We also find its variability to have the same period as its orbital period. \par 

X-ray luminosities are not provided by either the \textit{Chandra} and \textit{XMM-Newton} surveys for CPD--41 7717, so no conclusions can be made on its X-ray output. However, it has been established by the star's predicted companion masses that it is unlikely that the companion is a compact object, and so even if the system was an interacting one, the system should not be visible in the X-ray spectrum. \par

\begin{figure}
\centering
\includegraphics[width=\hsize]{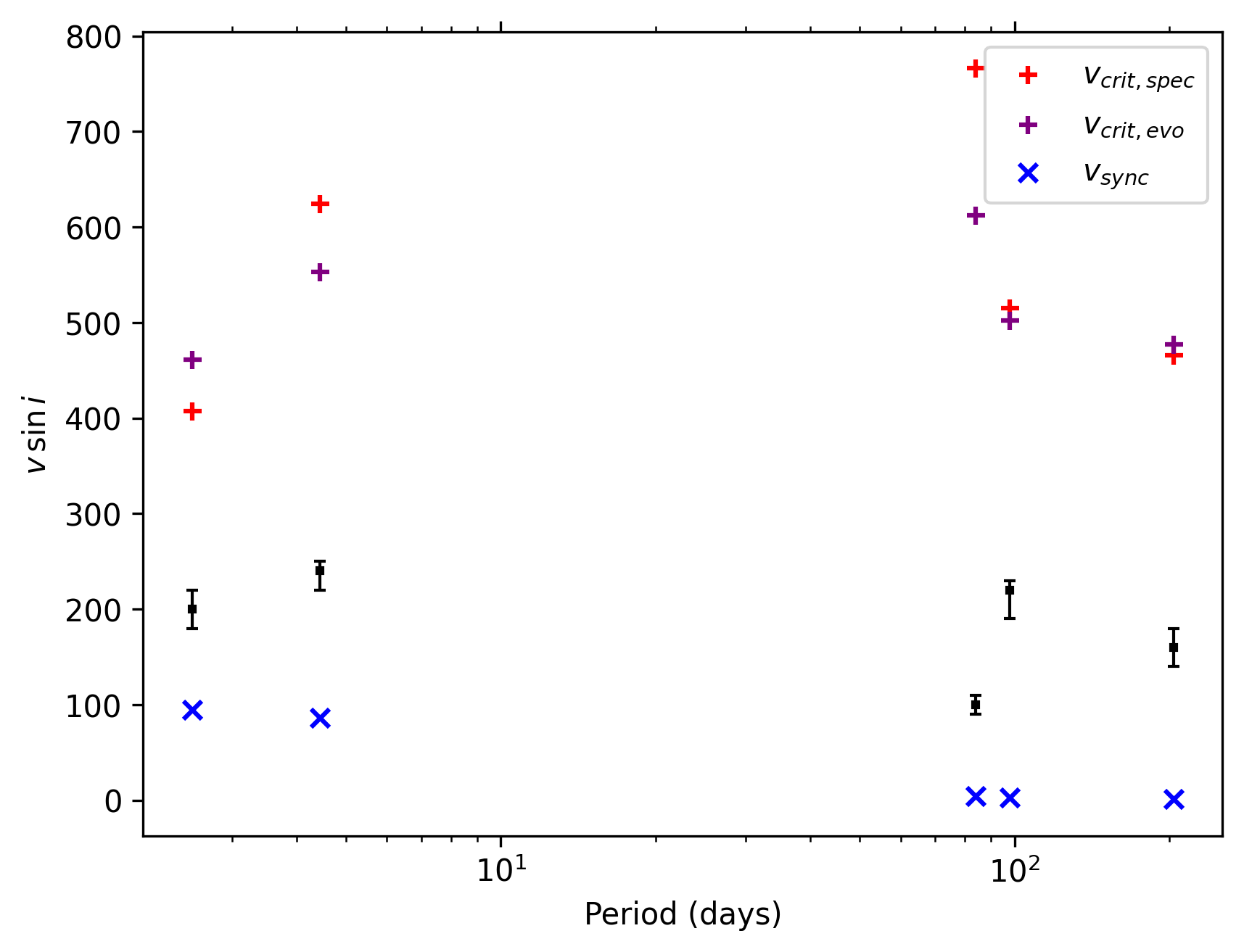}
    \caption{Orbital period against projected rotational velocity, $v\,\sin\,i$ (in black). Red plusses correspond to the critical rotational velocity, $v_{crit}$, calculated with Eqn. \ref{vcrit}. Blue crosses are the computed synchronous rotation rate, $v_{sync}$, computed for each star assuming they are in a synchronous orbit.}
    \label{f:vsinivsperiod}
\end{figure}

\subsection{Supersynchronous rotation}

Figure~\ref{f:vsinivsperiod} shows the constrained $v\,\sin\,i$ for the visible stars in candidate binaries with a compact object as a function of their orbital period. Also displayed here are the critical rotational velocities calculated for each star using Eqn. \ref{vcrit}, and using the parameters constrained with both the spectroscopic (and photometric) and evolutionary fitting. It is clear that none of the constrained values of $v\,\sin\,i$ are close to critical. \par 

However, for the shorter period systems (HD 152200 and CPD-41 7717), the constrained values of $v\,\sin\,i$ are around a factor of two higher than the theoretical synchronous rotation velocity, $v_{sync}$. This is assuming the rotational period of the star has tidally synchronised with the orbital period of its binary system, and so is computed with the orbital period of each system and the radius of the visible star. The synchronisation timescale, according to \cite{zahn_reprint_1977}, is proportional to $a^{\frac{17}{2}}$, where $a$ is the semi-major axis of the orbit. As a result, only close binary systems are expected to potentially exhibit this behaviour. \par 

To predict the synchronisation timescales of these shorter period systems and ascertain whether these two targets are compatible with them, we can follow what was done by \cite{el-badry_ngc_2022} for the proposed BH candidate NGC 2004 \#115 in the LMC \citep{lennon_vlt-flames_2021}. In that work, they use the Zahn theory, in which the predicted synchronization timescale is estimated to be  \par 

\begin{equation}
t_{\mathrm{sync}} = \frac{\beta}{5 \times 2^{5/3}E_{2}} \left( \frac{R^{3}}{GM} \right)^{1/2} \frac{1}{q^{2}(1+q)^{5/6}} \left( \frac{a}{R} \right)^{17/2}.
\label{tsync}
\end{equation}
where $M$ and $R$ are the mass and radius of the star undergoing synchronisation, in this case the primary star, $q$ is the mass ratio ($q = M_2/M$), $G$ is the gravitational constant, $a$ is the semi-major axis of the binary, $\beta = I/MR^2$, where $I$ is the primary star's moment of inertia, and $E_2$ is the tidal torque coefficient of the primary star. Further modelling of these systems, for example with MESA \citep{paxton_modules_2011}, is required to estimate $\beta$ and $E_2$. An analytic approximation for $E_2$, obtained by \cite{yoon_type_2010} from fitting the work of \cite{zahn_reprint_1977}, is $E_2 = 10^{-1.37}(R_{conv}/R)^{8}$, where $R_{conv}$ is the convective core radius. However, alternate analytic approximations for $E_2$, specifically for H-rich and He-rich stars have been formulated by \cite{qin_spin_2018}. They are $E_2 = 10^{-0.42}(R_{\mathrm{conv}}/R)^{7.5}$ for H-rich stars, and $E_2 = 10^{-0.93}(R_{\mathrm{conv}}/R)^{6.7}$ for He-rich stars. We consider each approximation for $E_2$ in our predictions of the synchronisation timescales of the two short-period systems.\par

For CPD-41 7717, we calculated a model for a 5 Myr-old 6.07 $M_{\odot}$ star using MESA (version 15140). The input physics of the simulations match those used in the Galactic \cite{brott_rotating_2011} models. We found that, from this model, $\beta = 0.06$, and $R_{conv}/R = 0.2$. Taking the average estimate for the unseen companion's mass from the spectroscopic mass estimate of the primary, we assume $M_2 = 0.65 M_{\odot}$. We also assume $a = 14 R_{\odot}$ from the system's period. If we use the same prescription \citep{yoon_type_2010} for $E_{2}$ used by \cite{el-badry_ngc_2022}, then we estimate the synchronisation timescale to be around 360 Myr. However, using the \cite{qin_spin_2018} prescriptions lead to a synchronisation timescale of around 15 Myr. Given this, it is unlikely for this system to be synchronised based on the cluster age.\par 

For HD 152200, a 5 Myr-old 23.35 $M_{\odot}$ MESA model was calculated. We found that, from this model, $\beta = 0.05$, and $R_{conv}/R = 0.2$. Like for CPD--41 7717, take the average spectroscopic estimate of HD 152200's unseen companion as the assumed secondary mass, $M_2 = 2.4 M_{\odot}$. We also assume $a = 34 R_{\odot}$ from the system's period. Using the \cite{yoon_type_2010} prescription for $E_{2}$ results in an estimated synchronisation timescale of around 60 Myr. However, using the \cite{qin_spin_2018} prescriptions lead to a a synchronisation timescale of around 3 Myr, which means that there is a possibility that the system should be synchronised based on what the assumption for the tidal torque coefficient is. \par

The potential overestimation of $v\,\sin\,i$ could be due to a few factors. The first is that we do not take the macroturbulent broadening of spectral lines into account during the spectroscopic fitting, as the tool we use does not do this. However, also have relatively low resolution spectra, and so we would have errors of the order of 10~\kms in constraining the macroturbelence. The second is that we have relatively low resolution spectra ($R\sim6500$). The third is that we have assumed that the rotation axis of these stars is perpendicular to the orbital plane. If the true $v\,\sin\,i$ is actually lower for these two close binaries, then the minimum inclination calculated with Eqn. \ref{inclination} will be smaller and as such, the lower bound of the range of unseen companion masses will increase, which decreases the likelihood of the unseen companion being a faint, non-degenerate star. \par

\section{Summary}
\label{s:summary}
In this work, we aimed to characterise the unseen companions of the 15 SB1 binary systems that were identified when studying the multiplicity of the B stars in NGC 6231 (in Paper~I), using multi-epoch medium-resolution optical spectroscopy, ground-based photometry and high-cadence space-based photometry to do so. Through Fourier disentangling of the spectra, we found 7 SB2 systems with mass ratios down to 0.1, extending past the limits of the original work in Paper~I. Out of the other 8 targets, 4 targets were found to be ambiguous cases where the nature of the extracted companion was unclear, and the other 4 targets had no signature of a faint companion in their spectra. Two of targets in the latter category were considered unlikely candidates for harbouring a compact object due to their late spectral type (B9) and their binary mass functions, and so were not considered for further study.\par 
Through atmospheric fitting, which combined both spectroscopic and photometric fitting against a TLUSTY model grid of synthetic spectra and SEDs, we estimated the spectroscopic mass and other stellar parameters of each visible primary star in the remaining SB1s, and then we provided a secondary estimate for the primary star masses through evolutionary fitting with BONNSAI. With these stellar parameters, and using the binary mass function already characterised for each SB1's orbit in Paper~I, we provide two estimates of the unseen companion's mass for our SB1 systems that we considered candidates for hosting a compact object companion. Consequently, we found four systems (HD 152200, V* V956 Sco, CD-41 11038, and CXOU J165421.3-415536) that have estimated companion masses that fall in the predicted mass ranges for Galactic compact objects, with the most likely BH host being CD-41 11038. However, the detection limits of our methodology do mean that it is possible for some or all of these candidates to have lower mass MS companions, stripped star companions or be part of triple systems.  \par 
We also looked at high-cadence TESS photometry to search for signatures of compact object companions in the light curves of these targets. However, as the field around these targets is very crowded, light curve extraction is challenging and we were unable to extract light curves for 6 out of the original 15 targets. Two of the targets showed eclipses in their light curve (V* V1208 and CPD-41 7746), and these two targets were confirmed to be newly disentangled SB2s. Three of the compact object hosting candidates have photometric variability in their lightcurves, though the origin of this variability is uncertain, and their periodicities do not appear to be consistent with signatures of compact object companions such as ellipsoidal variability and eclipsing. \par
We also find that the two close binaries out of the SB1s, HD 152200 and CPD--41 7717, appear to rotate supersynchronously, i.e. rotate above the speed at which the star would rotate if it was tidally synchronised with its companion. We estimated the tidal synchronisation timescales using \cite{zahn_reprint_1977} theory and MESA models, and we find that while it is not expected for CPD-41 7717 to be synchronised given the age of the cluster, it could be expected for HD 152200 to be. \par
The results presented in this work present four candidate SB1 systems for harbouring compact objects. Further photometric and interferometric follow-up on these candidates will help unambiguously characterise these systems, and consequently will potentially allow us to reduce the uncertainty in compact object formation scenarios. This work also presents a look at the more extreme end of the mass ratio distribution of these B-type stars, which is important input into the synthesis of populations of binary stars and studies of binary star formation.
%The authors thank the anonymous referee for the helpful comments, and constructive remarks on this manuscript. 

\begin{acknowledgements}
 
The research leading to these results has received funding from the European Research Council (ERC) under the European Union's Horizon 2020 research and innovation programme (grant agreement numbers 772225: MULTIPLES). L.M.\ thanks the European Space Agency (ESA) and the Belgian Federal Science Policy Office (BELSPO) for their support in the framework of the PRODEX Programme. H.S.\ acknowledges support from the FWO\_Odysseus program under project G0F8H6N. J.B. is supported by an ESO fellowship. T.S. acknowledges support from the European Union’s Horizon 2020 under the Marie Skłodowska-Curie grant agreement No 101024605. 

The TESS data presented in this paper were obtained from the Mikulski Archive for Space Telescopes (MAST) at the Space Telescope Science Institute (STScI), which is operated by the Association of Universities for Research in Astronomy, Inc., under NASA contract NAS5-26555. Support to MAST for these data is provided by the NASA Office of Space Science via grant NAG5-7584 and by other grants and contracts. Funding for the TESS mission is provided by the NASA Explorer Program. This research has made use of the SIMBAD database, operated at CDS, Strasbourg, France and of NASA's Astrophysics Data System Bibliographic Services. This research also made use of Lightkurve, a Python package for Kepler and TESS data analysis. The data used in this article will be shared on reasonable request to the corresponding authors.
%J.B.\ acknowledges support from the FWO Odysseus program under project G0F8H6N. J.I.V.\ acknowledges support from CONICYT-Becas Chile, ``Doctorado en el extranjero'' programme, Grant No. 72170619. This work has made use of data from the European Space Agency (ESA) mission {\it Gaia} (\url{https://www.cosmos.esa.int/gaia}), processed by the {\it Gaia} Data Processing and Analysis Consortium (DPAC, \url{https://www.cosmos.esa.int/web/gaia/dpac/consortium}). Funding for the DPAC has been provided by national institutions, in particular the institutions participating in the {\it Gaia} Multilateral Agreement. This publication makes use of data products from the Two Micron All Sky Survey, which is a joint project of the University of Massachusetts and the Infrared Processing and Analysis Center/California Institute of Technology, funded by the National Aeronautics and Space Administration and the National Science Foundation.
\end{acknowledgements}

\bibliographystyle{aa} % style aa.bst
\bibliography{paper2ref} % your references Yourfile.bib

\appendix

\section{Target spectra}
\label{app:spectra}

\begin{figure*}
\centering
\includegraphics[width=0.9\textwidth]{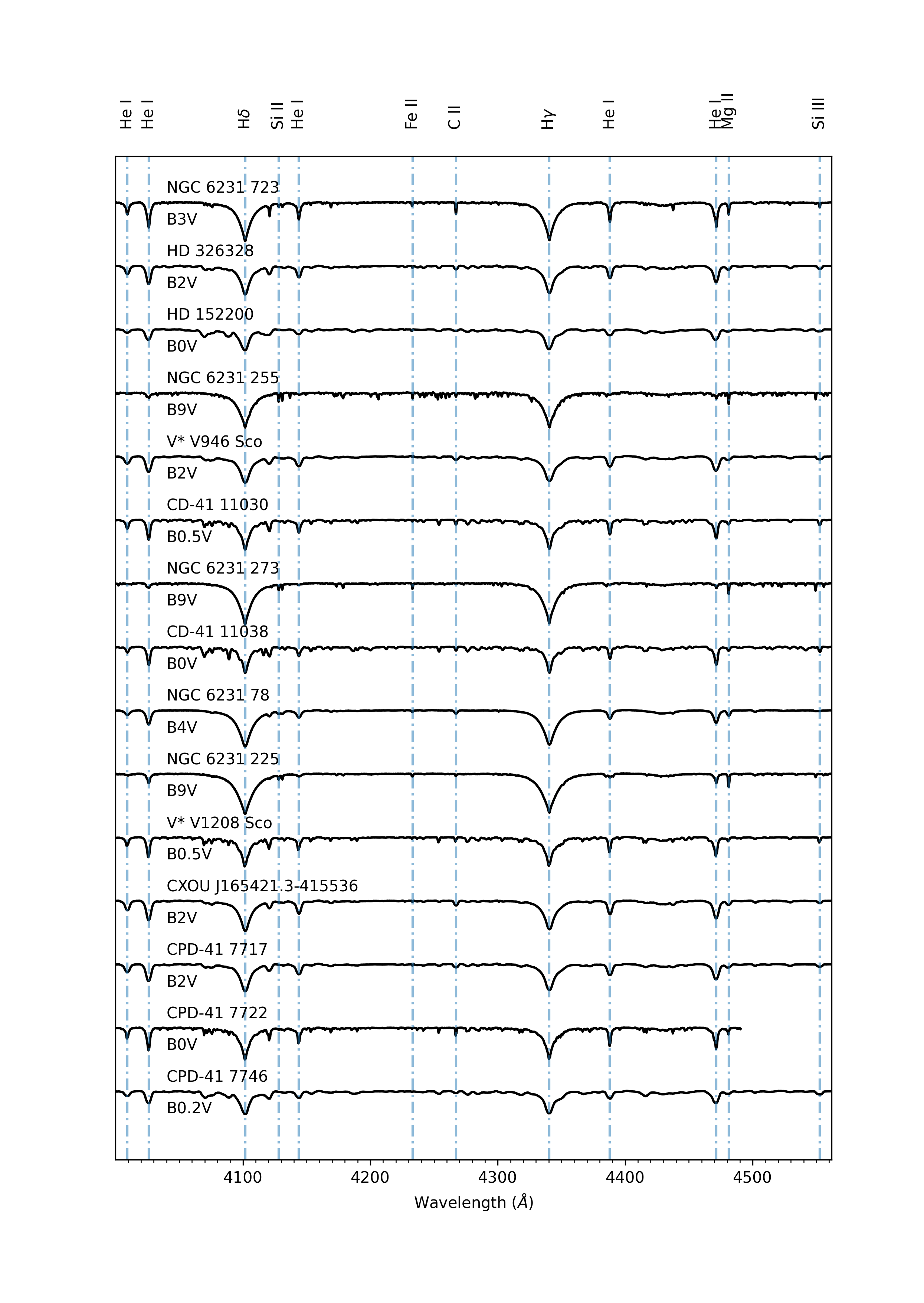}
    \caption{Shift-and-added VLT/FLAMES spectra for the targets (Table \ref{table:SB1}).
         }
    \label{FigSB1Spec}
\end{figure*}

\begin{figure*}
\centering
\includegraphics[width=0.9\textwidth]{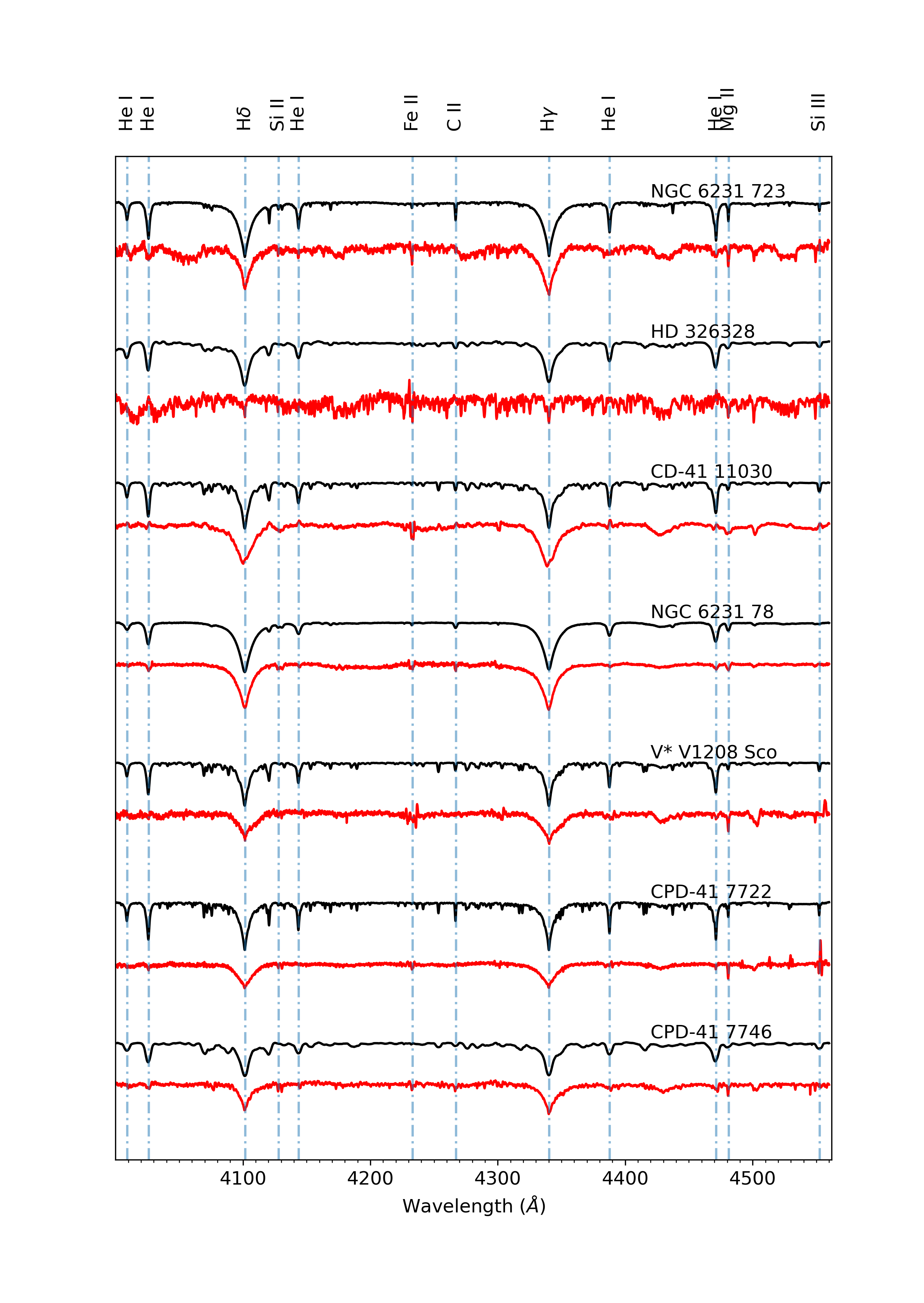}
    \caption{Disentangled VLT/FLAMES spectra for the newly classified SB2 systems in NGC 6231 (Section \ref{ss:disentangling})}.
    \label{FigSB2Spec}
\end{figure*}

\end{document}